\newcommand{\PaperTitle}{Exploring Collaborative Immersive Visualization \& Analytics for High-Dimensional Scientific Data through Domain Expert Perspectives}

\newcommand{\revise}[1]{\textcolor{black}{#1}}
\newcommand{\camera}[1]{\textcolor{black}{#1}}

\documentclass[sigconf]{acmart}

\AtBeginDocument{%
  }

\copyrightyear{2026}
\acmYear{2026}
\setcopyright{cc}
\setcctype{by}
\acmConference[CHI '26]{Proceedings of the 2026 CHI Conference on Human Factors in Computing Systems}{April 13--17, 2026}{Barcelona, Spain}
\acmBooktitle{Proceedings of the 2026 CHI Conference on Human Factors in Computing Systems (CHI '26), April 13--17, 2026, Barcelona, Spain}
\acmDOI{10.1145/3772318.3791203}
\acmISBN{979-8-4007-2278-3/2026/04}

\usepackage{graphicx}
\usepackage{xspace}
\usepackage{multirow}
\usepackage{enumitem}
\usepackage{hyperref}
\usepackage{fontawesome5}

\makeatletter
\newcommand{\quoting}{\@ifnextchar[{\quoting@opt}{\quoting@noopt}}
\newcommand{\quoting@opt}[2][]{%
  {\emph{``#2''}}~{\emph{[#1]}}%
}
\newcommand{\quoting@noopt}[1]{%
  {\emph{``#1''}}%
}
\makeatother

\begin{document}

\newboolean{showcomments}
\setboolean{showcomments}{true} % <- change to false to hide comments

% Fahim
\newcounter{fancounter}
\newcommand{\fan}[1]{%
  \ifthenelse{\boolean{showcomments}}{%
    \stepcounter{fancounter}%
    \textcolor{red}{\textbf{[Fahim \arabic{fancounter}:} #1]}%
  }{}%
}

% Reviewer
\newcounter{reviewcounter}
\newcommand{\review}[1]{%
  \ifthenelse{\boolean{showcomments}}{%
    \stepcounter{reviewcounter}%
    \textcolor{orange}{\textbf{[Review \arabic{reviewcounter}:} #1]}%
  }{}%
}

% Bo
\newcounter{cbhcounter}
\newcommand{\bo}[1]{%
  \ifthenelse{\boolean{showcomments}}{%
    \stepcounter{cbhcounter}%
    \textcolor{blue}{\textbf{[Han \arabic{cbhcounter}:} #1]}%
  }{}%
}

% Chen
\newcounter{csqcounter}
\newcommand{\csq}[1]{%
  \ifthenelse{\boolean{showcomments}}{%
    \stepcounter{csqcounter}%
    \textcolor{red}{\textbf{[Songqing Chen \arabic{csqcounter}:} #1]}%
  }{}%
}

% Li
\newcounter{licounter}
\newcommand{\jli}[1]{%
  \ifthenelse{\boolean{showcomments}}{%
    \stepcounter{licounter}%
    \textcolor{green!60!black}{\textbf{[Li \arabic{licounter}:} #1]}%
  }{}%
}

\title[]{\PaperTitle}

\author{Fahim Arsad Nafis}
\orcid{0009-0005-5803-5200}
\affiliation{
   \institution{George Mason University}
   \city{Fairfax}
   \state{Virginia}
   \country{USA}
}
\email{fnafis2@gmu.edu}

\author{Jie Li}
\orcid{0000-0002-6791-104X}
\affiliation{
   \institution{MIT Media Lab, MIT}
   \city{Boston}
   \state{Massachusetts}
   \country{USA}
 }
\email{jieli8@mit.edu}

\author{Simon Su}
\orcid{0000-0002-2460-3899}
\affiliation{
   \institution{National Institute of Standards and Technology}
   \city{Gaithersburg}
   \state{Maryland}
   \country{USA}
 }
\email{simon.su@nist.gov}

\author{Songqing Chen}
\orcid{0000-0003-4650-7125}
\affiliation{
   \institution{George Mason University}
   \city{Fairfax}
   \state{Virginia}
   \country{USA}
 }
\email{sqchen@gmu.edu}

\author{Bo Han}
\orcid{0000-0001-7042-3322}
\affiliation{%
   \institution{George Mason University}
   \city{Fairfax}
   \state{Virginia}
   \country{USA}
 }
\email{bohan@gmu.edu}

\renewcommand{\shortauthors}{Fahim Arsad et al.}

\begin{abstract}
Cross-disciplinary teams increasingly work with high-dimensional scientific datasets, yet fragmented toolchains and limited support for shared exploration hinder collaboration. Prior immersive visualization \& analytics research has emphasized individual interaction, leaving open how multi-user collaboration can be supported at scale. To fill this critical gap, we conduct semi-structured interviews with 20 domain experts from diverse academic, government, and industry backgrounds. Using deductive–inductive hybrid thematic analysis, we identify four collaboration-focused themes: workflow challenges, adoption perceptions, prospective features, and anticipated usability and ethical risks. These findings show how current ecosystems disrupt coordination and shared understanding, while highlighting opportunities for effective multi-user engagement. Our study contributes empirical insights into collaboration practices for high-dimensional scientific data visualization \& analysis, offering design implications to enhance coordination, mutual awareness, and equitable participation in next-generation collaborative immersive platforms. These contributions point toward future environments enabling distributed, cross-device teamwork on high-dimensional \camera{scientific} data.
\end{abstract}

\begin{CCSXML}
<ccs2012>
   <concept>
       <concept_id>10003120.10003130.10011762</concept_id>
       <concept_desc>Human-centered computing~Empirical studies in collaborative and social computing</concept_desc>
       <concept_significance>500</concept_significance>
       </concept>
   <concept>
       <concept_id>10003120.10003121.10011748</concept_id>
       <concept_desc>Human-centered computing~Empirical studies in HCI</concept_desc>
       <concept_significance>300</concept_significance>
       </concept>
 </ccs2012>
\end{CCSXML}

\ccsdesc[500]{Human-centered computing~Empirical studies in collaborative and social computing}
\ccsdesc[300]{Human-centered computing~Empirical studies in HCI}

\keywords{Immersive Analytics, Collaborative Systems, Scientific Visualization, High-Dimensional Data.}

\begin{teaserfigure}
  \centering
\includegraphics[width=\textwidth]{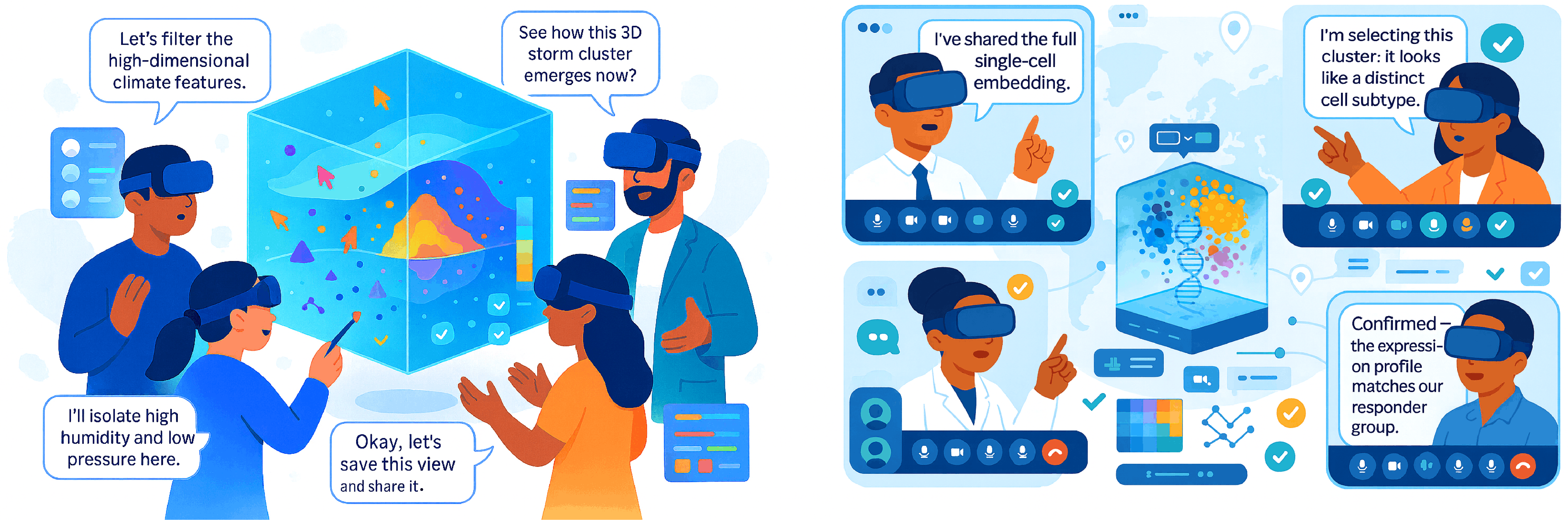}
  \caption{\revise{Conceptual overview illustrating synthesized scenarios of how domain scientists might use collaborative immersive visualization and analytics (CIVA). The figure presents two scenarios that integrate our analysis and design recommendations, grounded in participants’ accounts: (left) co-located collaborators jointly exploring high-dimensional climate simulation data in an immersive environment, and (right) geo-distributed collaborators coordinating around a shared single-cell embedding in a remote CIVA session.} Generated using GenAI tools and refined by the authors.}
  \Description{The left side of Figure 1 illustrates a co-located team of domain scientists using a collaborative immersive visualization and analytics (CIVA) system. A small group of collaborators wearing VR headsets stand in a semi-circle around a large holographic cube that encodes high-dimensional climate simulation data. Inside the cube, layered surfaces, colored clusters, and arrow-like vectors suggest multiple variables such as temperature, pressure, humidity, and wind, emphasizing that the system is designed for complex, multi-dimensional scientific data rather than simple charts. Subtle highlights, small floating interface elements, and pointing gestures indicate that the collaborators are jointly selecting, filtering, and interpreting regions of interest in the shared immersive space. The right side of Figure 1 illustrates a remote, geo-distributed CIVA scenario. Here, collaborators are shown in separate locations, each wearing a VR headset and positioned within a simplified local environment, conveying that they are connected through the system rather than physically co-present. In the center, a shared holographic data object encodes a high-dimensional single-cell embedding, rendered as dense point clouds and clustered structures reminiscent of dimensionality reduction outputs. Presence indicators, colored telepointer-style markers, and annotation icons highlight how remote participants can see each other’s focus, selections, and notes, supporting awareness and coordination around the same complex dataset despite being distributed across sites. Together, the two panels present synthesized scenarios that illustrate how domain scientists might use CIVA systems for co-located and remote collaboration with high-dimensional scientific data. They foreground recurring elements from our analysis—shared immersive workspaces, multi-dimensional climate and biological data, co-presence and presence awareness, and collaborative selection and annotation of regions of interest. By contrasting climate simulation exploration with single-cell embedding analysis, the figure conveys the cross-domain, multi-site nature of the use cases we target and visually motivates the design space and implications developed in the paper.}
  \label{fig: teaser}
\end{teaserfigure}

\maketitle

%%%%%%%%%%%%%%%%%%%%%%
\section{Introduction} 
%%%%%%%%%%%%%%%%%%%%%%
High-dimensional scientific datasets introduce challenges that extend beyond data volume, affecting domains such as climate modeling~\cite{johnstone2009statistical,holden2015emulation}, genomics~\cite{clarke2008properties,bermingham2015application,xing2001feature}, environmental physics~\cite{turmon2010statistical,su2016multi}, high-performance computing~\cite{exposito2013performance,benedict2013performance}, artificial intelligence~\cite{aziz2018artificial,jain2023artificial}, and quantum computing~\cite{peters2021machine,zhao2024unraveling,Shafique2024Quantum}. These datasets exhibit heterogeneity, temporal dynamics, and inter-dependencies that conventional methods struggle to visualize, as seen in multivariate simulations~\cite{tang2016visualizing,wright2022high}, protein folding~\cite{itoh2009multidimensional}, and quantum entanglement tasks~\cite{ma2024high}. For domain experts, these complexities become most acute during interpretation, where clusters of varying shapes, sizes, and densities must be identified amid noise~\cite{ertoz2003finding,assent2012clustering}. Without careful evaluation, conventional reduction methods, such as PCA~\cite{abdi2010principal} and t-SNE~\cite{maaten2008visualizing}, risk over-simplification, distortion, and reduced trust in analytical outcomes~\cite{ahadzadeh2023sfe,wright2022high}. Interactive visualization is therefore crucial, allowing experts to query, compare, and explore multidimensional scientific data~\cite{buja1996interactive}.

Visualization platforms such as ParaView~\cite{paraview,paraview_web}, VisIt~\cite{visit}, Tecplot~\cite{tecplot}, and VAPOR~\cite{VAPOR} provide established pipelines for volumetric rendering, large-scale data processing, and temporal exploration. However, these systems remain tied to desktop-based interaction models, where projecting multivariate structures onto two-dimensional displays constrains spatial reasoning and limits fluid exploration~\cite{tang2016visualizing,engel2012survey,liu2016visualizing,johnstone2009statistical,inselberg2002visualization}. Immersive extensions such as ParaView’s \textit{extended reality (XR) interface}~\cite{paraview-xr-interface} have emerged, yet evaluations report shortcomings in synchronous manipulation, provenance capture, and handling real-time updates for time-evolving datasets~\cite{nafis2024we}. These gaps are particularly evident in distributed collaborations where domain scientists coordinate across institutions and often rely on screenshots, static renderings, or asynchronous reports~\cite{sun2023dynamic}. Consequently, existing workflows fragment communication and hinder interactive, real-time negotiation of insights that modern scientific analytics requires~\cite{buja1996interactive,bertini2011quality}.

Immersive analytics (IA)~\cite{marriott2018immersive,goncu2015immersive,skarbez2019immersive}, empowered by virtual, augmented, and mixed reality (VR/AR/MR), is increasingly regarded as a promising complement to established visualization ecosystems~\cite{fonnet2019survey}. It enables domain experts to interact with high-dimensional data beyond desktop limitations~\cite{keim2008visual,van2000immersive,laha2012identifying,andrews2010space}. In immersive environments, researchers can navigate volumetric datasets, inspect structures from multiple perspectives, and annotate features using six degrees of freedom (6DoF)~\cite{stewart1965platform}, while collaborators interact in real time~\cite{cordeil2017imaxes,prouzeau2019scaptics}. This opens possibilities for multi-user co-presence, joint manipulation of complex models, and contextual annotation of findings, which remain difficult to achieve with traditional tools~\cite{billinghurst2018collaborative,hackathorn2016immersive,marai2016interdisciplinary,qiu2022vicollar,cavallo2019dataspace}. Immersive applications are emerging across domains: to explore atmospheric models~\cite{zhao2019harnessing,helbig2014concept}, to analyze biomedical imagery~\cite{knodel2018virtual,czauderna2018immersive,prodromou2020multi}, and to support digital twin simulations for industrial processes and performance optimization~\cite{del2023digital,geringer2024mint}. Beyond spatial reasoning, these platforms promise closer integration with scientific cyberinfrastructure, creating opportunities for seamless collaboration across distributed teams~\cite{marriott2018immersive,ens2022immersive}. By lowering participation barriers
and supporting dynamic modes of interpretation, IA can reshape how scientific communities collectively explore complex datasets~\cite{goncu2015immersive}.

Despite this progress, a critical gap persists in understanding how IA can be meaningfully integrated into real-world collaborative scientific workflows~\cite{nguyen2019collaborative,vaslin2025informal,lee2020shared,borowski2025dashspace,srinivasan2025heedvision,hackathorn2016immersive,reski2022empirical,zagermann2023challenges}. Existing systems largely emphasize proof-of-concept prototypes~\cite{cordeil2019iatk,cordeil2017imaxes} or controlled evaluations with non-domain experts~\cite{goncu2015immersive,ens2021grand}, 
providing limited insight into the lived practices of scientists who analyze high-dimensional data across interdisciplinary, institutionally distributed teams. Collaborative immersive visualization and analytics (CIVA) remains particularly underexplored: a survey of IA research from 1991 to 2018 found that around 15 of 127 system papers explicitly addressed collaboration~\cite{fonnet2019survey,billinghurst2018collaborative}, despite scientists consistently identifying it as a priority. Reported limitations, such as difficulties in synchronous manipulation, communication constraints, absence of role-based access, and provenance challenges~\cite{nafis2024we,borowski2025dashspace,srinivasan2025heedvision,reski2022empirical,lee2020shared} reveal a persistent disconnect between the demands of scientific practice and the capabilities of existing immersive systems.

\camera{Beyond these technical and infrastructural gaps, existing work offers limited conceptual accounts of collaboration itself in IA settings. In CIVA environments, we find collaboration is organized around shared spatial–analytic states that collaborators jointly perceive, manipulate, and interpret. These interactions unfold through real-time coordination not fully captured by frameworks that treat collaboration as artifact-mediated coordination among individuals. Recognizing this distinction is essential for understanding both the limitations of current immersive systems and the forms of collaborative support that future CIVA platforms must provide.}

To address this \camera{conceptual and empirical} gap, we investigate how domain scientists organize analytical and visualization workflows and coordinate around high-dimensional data. We then surface the concrete challenges they face and the CIVA-supported solutions they envision. Understanding these practices is essential for designing \camera{and conceptualizing} CIVA systems that move beyond lab-scale prototypes and integrate into everyday scientific practice. Guided by this motivation, we investigate the following research questions:

\begin{itemize}[leftmargin=*]
    \item \textbf{RQ1:} \textit{How do domain scientists analyze, visualize, and collaborate around high-dimensional scientific data, and what limitations arise in their current tools and workflows?}
    \item \textbf{RQ2:} \textit{What expectations, concerns, and constraints do domain scientists anticipate when considering the adoption of CIVA systems to support their collaborative analysis and visualization workflows?}
    \item \textbf{RQ3:} \textit{What features and capabilities do domain scientists envision for future CIVA systems to better support their collaborative analysis and visualization workflows?}
\end{itemize}

To explore these questions, we conducted semi-structured interviews~\cite{adams2015conducting} lasting approximately one hour with domain scientists across various scientific fields. 
Interviews covered analytical workflows, collaboration practices, perspectives on immersive tools, responses to design probes through demonstration videos, and reflections on future needs. All interviews were audio-recorded, transcribed verbatim, and analyzed collaboratively in a shared digital workspace. We employed a hybrid thematic analysis combining deductive coding with inductive affinity diagramming. 

This paper provides an empirical account of how domain scientists work with high-dimensional data and envision the role of CIVA within scientific visualization workflows, as conceptually illustrated in Figure~\ref{fig: teaser}, which contrasts co-located teams with geographically distributed collaborators engaging in immersive exploration of complex datasets. Our contributions are: 
\begin{itemize}[leftmargin=*]
    \item {Characterization of current analytical and visualization workflows followed by domain scientists working with high-dimensional data.}
    \item {Identification of challenges in existing tools, workflows, and collaborative processes for high-dimensional scientific data analysis and visualization.}
    \item {Synthesis of domain experts’ expectations and concerns regarding the adoption of CIVA, including usability, performance, accessibility, cost, and ethical considerations.}
    \item {Derivation of prospective features and functional requirements envisioned for future CIVA systems.}
    \item {Design implications for next-generation CIVA platforms that translate these findings into actionable directions for supporting collaborative, high-dimensional data work.}
\end{itemize}

Collectively, these contributions lay the empirical groundwork for advancing CIVA toward practical deployment in real-world scientific workflows. \camera{The remainder of this paper is organized as follows: Section ~\ref{related_work} reviews related work; Section ~\ref{methodology} describes the study methodology; Section ~\ref{finding} presents the findings; Section ~\ref{implication} discusses design implications; Section ~\ref{limitation} outlines limitations and future work; and Section ~\ref{conclusion} concludes the paper.}

%%%%%%%%%%%%%%%%%%%%%%
\section{Background and Related Work}\label{related_work}
%%%%%%%%%%%%%%%%%%%%%%

\begin{figure*}[th]
  \centering
  \includegraphics[width=\textwidth]{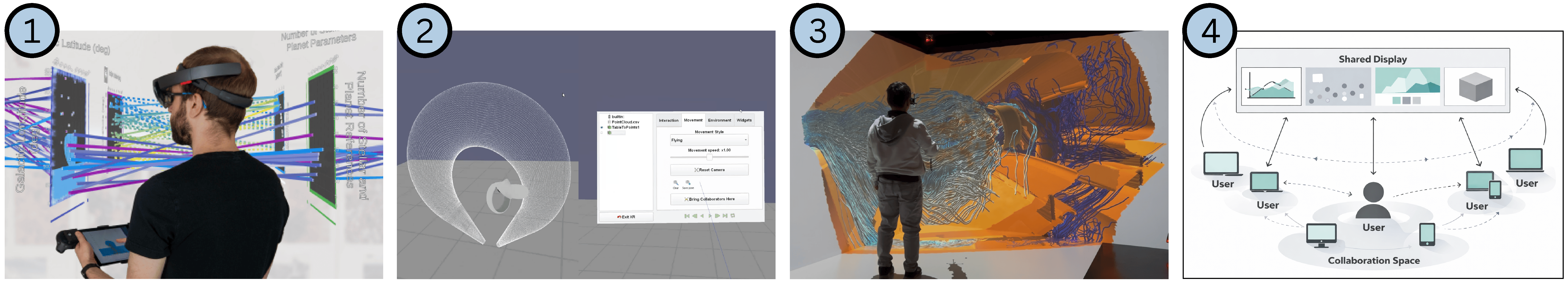}
  \caption{Examples of immersive visualization and analytics platforms across different modalities. (1) STREAM~\cite{hubenschmid2021stream} integrating spatially-aware tablets with AR head-mounted displays for multimodal 3D data interaction. \camera{\copyright~2021 ACM. Image reused under CC BY 4.0;}
  (2) ParaView’s XR Interface plugin used in VR with an Oculus Quest 2 headset for exploring high-dimensional scientific data; 
  (3) A CAVE installation at the National Institute of Standards and Technology (NIST) supporting room-scale immersion with HPC-driven datasets; and 
  (4) \camera{A conceptual illustration of a SAGE2~\cite{marrinan2014sage2} tiled display wall environment depicting collaborative use of images, videos, documents, and 2D/3D data applications across a shared display and personal devices. Generated using GenAI tools and refined by the authors.}
  }
  \Description{The figure shows a collage of four images representing immersive analytics platforms. 
  (a) STREAM integrates AR head-mounted displays with spatially-aware tablets, enabling multimodal interaction with colorful 3D data. 
  (b) A scientist wearing a VR headset (Oculus Quest 2) uses ParaView’s XR Interface plugin to analyze high-dimensional datasets. 
  (c) A CAVE installation at NIST surrounds users with wall-sized immersive displays of scientific data running on HPC systems. 
  (d) A conceptual illustration of a SAGE2 tiled display wall environment depicting collaborative use of images, videos, documents, and 2D/3D data applications across a shared display and personal devices. The illustration conveys how multiple users simultaneously engage with heterogeneous data sources through coordinated interaction between a large, high-resolution display wall and individual personal devices, supporting shared awareness and collaborative analysis.}
  \label{fig:immersive_platforms}
\end{figure*}

This section outlines the foundations of high-dimensional scientific data, immersive analytics, and collaborative immersive visualization and analytics, placing our study in the research context.

\subsection{High-dimensional Scientific Data}
Domains such as genomics, climate modeling, astronomy, and finance are experiencing rapid growth in high-dimensional data,
where the number of features \(p\) greatly exceeds the number of observations \(n\) (\(p \gg n\))~\cite{johnstone2009statistical, liu2016visualizing,engel2012survey}. This imbalance intensifies the \textit{curse of dimensionality}~\cite{verleysen2005curse}, leading to unreliable distances, unstable density estimates, and exponentially complex feature selection~\cite{li2017feature}.
Experts must also make sense of interwined multivariate, temporal, and spatial structures, as seen in evolving climate fields~\cite{bremer2010interactive,pascucci2010topological,tang2016visualizing} or biological cell trajectories~\cite{satija2015spatial,regev2017human,zhang2021spatially}. Dimensionality reduction techniques address these issues by projecting high-dimensional relations into interpretable spaces. Linear methods such as PCA~\cite{abdi2010principal} and MDS~\cite{cox2008multidimensional} preserve variance or distance, while nonlinear approaches, including t-SNE~\cite{maaten2008visualizing} and UMAP~\cite{mcinnes2018umap}, reveal neighborhood patterns and local clusters in complex datasets~\cite{liu2016visualizing}.
Visual encoding methods complement these projections through scatterplot matrices~\cite{wilkinson2005graph}, parallel coordinates~\cite{inselberg1985plane}, and glyph- or pixel-based representations~\cite{kandogan2000star,keim2002designing}. Interactive systems such as GGobi~\cite{swayne2003ggobi}, XmdvTool~\cite{ward1994xmdvtool}, VisDB~\cite{keim2002visdb} add brushing, linking, and filtering~\cite{tatu2009combining} but remain primarily single-user and constrained to desktop-bound.

Despite these advances, major barriers persist. Visual scalability breaks down as many encodings collapse into clutter when applied to millions of records~\cite{liu2016visualizing}. Interpretability is also limited; nonlinear embeddings can distort relationships~\cite{maaten2008visualizing,mcinnes2018umap}, axis-ordering alters patterns in parallel coordinates~\cite{lu2012clutter}, and dimensionality reduction often obscures semantic loss to domain variables~\cite{bertini2011quality}. Analysts must additionally manage high cognitive load when tuning parameters, comparing projections, and coordinating multiple linked views~\cite{swayne2003ggobi}. Widely used toolchains in Python (e.g., NumPy~\cite{numpy}, Pandas~\cite{pandas_web}, scikit-learn~\cite{sklearn}, and Plotly Dash~\cite{plotly_dash}), R (ggplot2~\cite{ggplot}), and scientific platforms (e.g., ParaView~\cite{paraview}, VTK~\cite{vtk}, VisIt~\cite{visit}, and QGIS~\cite{qgis}), provide robust capabilities but remain oriented toward single-user, desktop-bound workflows with limited support for multi-user interpretation. Recent surveys~\cite{engel2012survey,liu2016visualizing} underscore that achieving scalable, interpretable, and collaborative visualization of high-dimensional data remains an open challenge. These limitations have prompted growing interest in immersive approaches that use spatial interaction and embodied perception to support the exploration of complex, high-dimensional data.

%underscore that achieving scalable, interpretable, and collaborative visualization of high-dimensional data remains an open challenge, motivating emerging work on immersive and multi-user environments.

%%%%%%%%%%%%%%%%%%%%%%
\subsection{Immersive Analytics}
%%%%%%%%%%%%%%%%%%%%%%
Immersive analytics (IA) explores how augmented reality~\cite{azuma1997survey}, virtual reality~\cite{sutherland1968head}, CAVE environments~\cite{cruz2023surround}, and large display walls~\cite{andrews2010space}
transform the analysis of high-dimensional and spatially complex data
~\cite{goncu2015immersive,dwyer2018immersive,bach2016immersive}, as presented in Figure~\ref{fig:immersive_platforms}. Positioned at the intersection of visualization, HCI, and extended reality, IA research 
articulates theoretical frameworks and design principles for immersive perception, interaction, and workflow integration~\cite{marriott2018immersive,fonnet2019survey,skarbez2019immersive,ens2021grand}. Key affordances  
include depth-enhanced perception through stereoscopy and motion parallax~\cite{ware2005reevaluating,marriott2018immersive}, spatial navigation for multivairate exploration~\cite{ens2022immersive}, and multimodal input via gesture, gaze, haptics, and voice~\cite{buschel2018interaction,yang2022towards}. These support externalizing cognition into 3D workspaces, improving pattern recognition and sensemaking~\cite{elsayed2016situated}. Toolkits and platforms such as IATK~\cite{cordeil2019iatk}, DXR~\cite{sicat2018dxr}, immersive extensions of ParaView~\cite{paraview} and Catalyst~\cite{catalyst}, and systems such as SlicerVR~\cite{pinter2020slicervr}, Nanome~\cite{nanome}, and VisNEST~\cite{nowke2013visnest} demonstrate multidimensional exploration and situated analysis across biomedical, geospatial, urban, and neural data domains~\cite{westwood2005immersive, zhao2019harnessing,xu2023toward}.

However, IA continues to face practical and technical constraints. Physiological issues such as simulator sickness and the vergence accommodation conflict reduce comfort and sustained use~\cite{marriott2018immersive,ens2021grand}. Cost, setup complexity, and physical space requirements create practical barriers~\cite{ens2022immersive}, while technical challenges arise from rendering large datasets, unstable pipelines, and the lack of standardized interaction paradigms~\cite{fonnet2019survey}.
Evaluation practices also remain fragmented, prioritizing short-term task accuracy over long-term workflow integration or collaboration~\cite{friedl2024systematic,skarbez2019immersive}. These limitations highlight the need for scalable, multi-user, and workflow-aware immersive infrastructures, an area that has been repeatedly emphasized yet remains underexplored in practice.

%Together, these constraints indicate that while IA has matured as an interaction paradigm, its integration into sustained, collaborative scientific workflows remains limited. This gap underscores the need for scalable, multi-user, and workflow-aware immersive infrastructures, an area that has been repeatedly emphasized yet remains underexplored in practice.

%%%%%%%%%%%%%%%%%%%%%%
\subsection{Collaborative Immersive Visualization and Analytics}
%%%%%%%%%%%%%%%%%%%%%%

Extending immersive visualization and analytics into multi-user contexts enables participants to jointly explore and interpret data within shared AR/VR environments, which is the defining aim of collaborative immersive visualization and analytics (CIVA)~\cite{billinghurst2018collaborative,hackathorn2016immersive}. Prior work emphasizes how spatial co-presence, synchronized viewpoints, and embodied interaction support joint sensemaking in high-dimensional settings~\cite{benk2022my}. Co-located systems such as FIESTA~\cite{lee2020shared} demonstrate group exploration on shared immersive surfaces, while distributed platforms such as DashSpace~\cite{borowski2025dashspace} use WebXR to coordinate avatars, annotations, and cross-toolkit visualizations. Scientific domains support multi-user XR tools including ParaView XR~\cite{paraview-xr-interface}, provenance-enabled VisIt extensions~\cite{visit,nguyen2016collaborative}, SlicerVR~\cite{pinter2020slicervr}, and embodied multidimensional environments such as ImAxes~\cite{cordeil2017imaxes}. Hybrid systems, such as  SAGE3~\cite{harden2023sage3} and VRIA~\cite{butcher2020vria}, bridge immersive and desktop collaboration with web-based provenance support.

Recent multi-user XR systems broaden this design space. For example, eyemR-Vis~\cite{jing2021eyemr} visualizes bi-directional gaze cues for joint attention. Volumetric telepresence frameworks, such as Virtual Co-Presence~\cite{irlitti2023volumetric} and Volumetric Hybrid Workspaces~\cite{irlitti2024volumetric}, and real-time point-cloud streaming~\cite{lee2021sharing} enable shared spatial context across sites. Edge-based platforms, including ARENA~\cite{pereira2021arena}, WiXaRd~\cite{stacchio2024wixard}, and SCAXR~\cite{huang2023scaxr}, provide scalable synchronization, provenance management, and heterogeneous device support, demonstrating rapidly evolving capabilities for distributed and co-located immersive collaboration.

Despite these developments, CIVA faces substantial challenges. Multi-user synchronization must remain low-latency for shared scene states, embodiment, and annotations~\cite{zagermann2023challenges}, and coordination issues persist even with cues such as gaze or field-of-view indicators~\cite{srinivasan2025heedvision}. Accessibility barriers, including hardware cost, ergonomic strain, and spatial constraints, further limit adoption, disproportionately affecting disabled users~\cite{creed2024inclusive, reski2022empirical}. Technical hurdles arise from the computational demands of large scientific datasets, unstable XR pipelines, and the lack of standardized cross-device interaction models, and challenges in supporting geographically dispersed collaboration.~\cite{fonnet2019survey}. Beyond technical constraints, prior systems often embed implicit assumptions about roles, expertise, and collaboration structure, yet rarely examine how domain experts actually coordinate, negotiate meaning, and distribute work within IA settings. Moreover, most evaluations remain short-term and prototype-focused~\cite{ens2021grand,hackathorn2016immersive}, leaving open questions about long-term workflow integration, multi-user provenance, and expert collaboration. These gaps motivate scalable, accessible CIVA infrastructures grounded in domain experts’ real practices.
%%%%%%%%%%%%%%%%%%%%%%
\section{Methodology} \label{methodology}
%%%%%%%%%%%%%%%%%%%%%%
\camera{This section describes the overall study approach, including how participants were recruited, how data were collected through interviews, and how the resulting data were analyzed.}
\begin{table*}[t]
\centering
\small
\begin{tabular}{l|l|p{8cm}}
\hline
%\rowcolor{black!20}
\multicolumn{3}{|c|}{\textbf{Specialized Research Domains}} \\
\hline
\multicolumn{2}{l|}{\textbf{Characteristics}} & \textbf{Participants} \\
\hline
\multicolumn{2}{l|}{Cross-cutting Methods \& Data Modalities} & P1, P12, P13 \\
\multicolumn{2}{l|}{Engineering \& Computational Infrastructure} & P2, P18, P20 \\
\multicolumn{2}{l|}{Physical \& Space Sciences} & P3 \\
\multicolumn{2}{l|}{Earth \& Environmental Systems \& Policy} & P4, P6, P8, P11, P14, P19 \\
\multicolumn{2}{l|}{Medical, Health \& Neuroscience} & P5, P7, P16 \\
\multicolumn{2}{l|}{Finance \& Banking Analytics} & P9, P10 \\
\multicolumn{2}{l|}{Sports \& Human Performance Analytics} & P15, P17 \\
\hline
%\rowcolor{black!20}
\multicolumn{3}{|c|}{\textbf{Familiarity with Data Types}} \\
\hline
\textbf{Characteristics} & \textbf{Data Type} & \textbf{Participants} \\
\hline
\multirow{3}{*}{Spatial \& Environmental} 
& LiDAR & P2, P6, P16, P18 \\
& Satellite imagery & P3, P4, P8, P11, P13, P16 \\
& Climate models & P4, P6, P11, P16, P19 \\
\hline
\multirow{3}{*}{Biological \& Medical} 
& Medical imaging & P2, P5, P6, P14, P16 \\
& Genomic sequences & P16 \\
& Medical and health care data & P7 \\
\hline
\multirow{4}{*}{Temporal \& Sequential} 
& Time-series data & P3, P4, P5, P6, P7, P8, P9, P10, P11, P12, P13, P14, P15, P16, P17, P18, P20 \\
& Video data & P1, P3, P6, P13, P15, P17 \\
& Audio data & P6, P8, P12, P13 \\
& Qualitative / perceptions & P19 \\
\hline
\multirow{2}{*}{3D \& Sensor} 
& 3D point clouds / meshes & P2, P3, P5, P6, P13, P16, P18 \\
& Robotics sensor data & P2, P13, P17 \\
\hline
Multimodal / Integrated 
&  Combining text, image, audio & P2, P6, P8, P9, P11, P13, P16, P18, P20 \\
\hline
%\rowcolor{black!20}
\multicolumn{3}{|c|}{\textbf{Primary Tools for Data Analysis and Visualization}} \\
\hline
\textbf{Characteristics} & \textbf{Tools / Platforms} & \textbf{Participants} \\
\hline
\multirow{3}{*}{Programming Language} 
& Python (matplotlib, seaborn, etc.) & P1, P2, P4, P5, P6, P7, P10, P12, P13, P14, P16, P17, P18, P20 \\
& R (ggplot2, Shiny, etc.) & P1, P4, P5, P8, P11, P16, P17, P19 \\
& MATLAB & P1, P16, P17 \\
\hline
\multirow{2}{*}{Visualization Platform} 
& ParaView & P2, P3, P5, P6, P20 \\
& IDL & P3 \\
& AVS & P20 \\
\hline
Statistical Software & STATA & P7 \\
\hline
\multirow{2}{*}{Business Intelligence Tool} 
& Tableau & P9, P10 \\
& PowerBI & P11 \\
\hline
Web-based Visualization & D3 and Java & P15 \\
\hline
Immersive Visualization & Unreal Engine, NVIDIA Omniverse & P18 \\
\hline
Cloud Computing 
& Google Colab, Azure DevOps, AWS
& P1, P2, P4, P7, P9, P10, P12, P13, P14, P18 \\
\hline
%\rowcolor{black!20}
\multicolumn{3}{|c|}{\textbf{Collaborative Software Usage}} \\
\hline
\textbf{Characteristics} & \textbf{Tools / Platforms} & \textbf{Participants} \\
\hline
Communication 
& Slack, Microsoft Teams, Discord, Zoom 
& P1, P2, P4, P5, P8, P9, P10, P11, P12, P13, P15, P16, P17, P18, P20 \\
\hline
Project Management 
& Trello, Asana, Jira, Notion 
& P1, P9, P17 \\
\hline
Version Control
& GitHub, GitLab, Bitbucket, SourceForge 
& P1, P3, P4, P5, P6, P9, P10, P11, P12, P13, P14, P15, P16, P17, P18, P20 \\
\hline
Storage \& Sharing 
& Google Drive, OneDrive, Dropbox
& P1, P2, P3, P5, P6, P7, P8, P9, P10, P11, P12, P13, P15, P16, P17, P18, P19 \\
\hline
\end{tabular}
\caption{Domains, datasets, visualization tools, and collaboration platforms categorized with participant identifiers.}
\label{tab: participant details}
\end{table*}

%%%%%%%%%%%%%%%%%%%%%%
\subsection{Study Design and Recruitment} 
%%%%%%%%%%%%%%%%%%%%%%

We conducted a qualitative study using semi-structured interviews~\cite{adams2015conducting} with domain experts who routinely work with high-dimensional scientific data using purposive sampling~\cite{tongco2007purposive} to capture variation in disciplinary areas, professional experience, organization type, and geographic location across multiple continents. Recruitment proceeded via a screening survey distributed through professional networks (e.g., LinkedIn), specialist forums, and mailing lists in data analytics, visualization, and IA communities. The survey captured respondents’ roles, scientific domain, analytic and collaborative tools, and familiarity with immersive technologies. It was used to assess inclusion criteria: participants had to (a) currently work with high-dimensional scientific data, (b) have experience with data analysis or visualization tools, and (c) be at least $18$ and able to complete an online interview in English; prior immersive-technology experience was not required. $48$ respondents met these criteria, from which we purposively selected $20$ to ensure balance, and contacted them individually to schedule interviews. All 20 participants completed individual online interviews lasting 60–75 minutes between March and July 2025. Each was offered a 20 USD (or local equivalent) gift card upon completion; 4 declined compensation. This sample size is considered sufficient, as thematic saturation in qualitative studies is often reached around 15–17 interviews, and aligns with reported averages for remote HCI interviews ($M=15, SD=6$)~\cite{caine2016local}. The first author led all interviews, which were audio-recorded and transcribed. Recruitment and data collection followed IRB requirements where applicable and complied with the lead author’s institutional ethics protocol, including secure storage of audio recordings and de-identified transcripts.

\begin{figure*}[t]
  \centering
  \includegraphics[width=\textwidth]{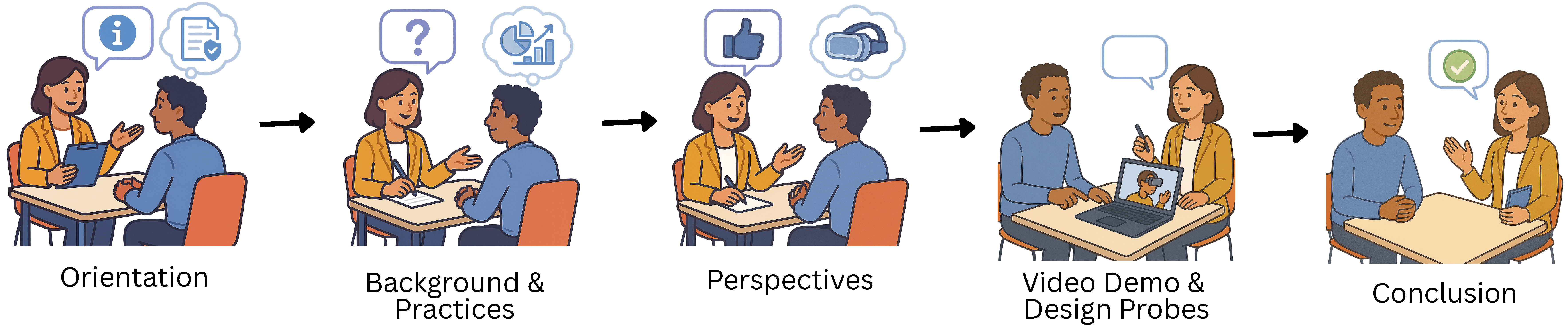}
  \caption{Sequential stages of the user interview process. Generated using GenAI tools and refined by the authors.}
  \Description{This figure illustrates the structured flow of a user interview study, depicted through five sequential stages. In the \emph{Orientation} stage, the interviewer introduces the study, clarifies objectives, and ensures consent. The \emph{Background \& Practices} stage explores participants’ existing roles, workflows, tools, and collaboration challenges. In the \emph{Perspectives} stage, participants share impressions of potential benefits, risks, and suitability of immersive environments for analytical tasks. The \emph{Video Demo \& Design Probes} stage engages participants in reflecting on a demonstration of immersive analytics, prompting discussion of feature fit, hurdles, mitigations, and ethical considerations. Finally, the \emph{Conclusion} stage allows participants to summarize their views and offer final recommendations for improving collaboration features. Across all panels, consistent characters and visual cues highlight structured yet human-centered research interactions.}
  \label{fig:interview_stages}
\end{figure*}

%%%%%%%%%%%%%%%%%%%%%%
\subsection{Participants}
%%%%%%%%%%%%%%%%%%%%%%

To ensure traceability in presenting the results, we labeled the participants as P1–P20. The participants reflected diversity across age, race, educational background, and professional roles. The cohort comprised 16 male and 4 female participants across age groups: under 25 $(n=1)$, 25–34 $(n=14)$, 35–44 $(n=1)$, 45–54 $(n=1)$, and 55–64 $(n=3)$. Participants represented racial backgrounds: \textit{South Asian} $(n=6)$, \textit{East Asian} $(n=5)$, \textit{White} $(n=4)$, \textit{Hispanic/Latino} $(n=2)$, \textit{Southeast Asian} $(n=1)$, \textit{Multiracial} $(n=1)$, and one who preferred not to disclose. Educational backgrounds included \textit{doctoral} $(n=13)$, \textit{master’s} $(n=4)$, and \textit{bachelor’s} $(n=3)$ degrees. Professional roles included \textit{researchers} $(n=8)$, \textit{engineers} $(n=4)$, \textit{graduate students} $(n=4)$, and \textit{other related roles} $(n=4)$, distributed across \textit{academic institutions} $(n=11)$, \textit{public-sector agencies} $(n=4)$, \textit{private companies} $(n=3)$, \textit{non-profit organization} $(n=1)$, and \textit{healthcare institution} $(n=1)$. Experience in their primary field ranged from 1–3 years $(n=8)$, 4–5 years $(n=4)$, and 6–10 years $(n=2)$, to more than 10 years $(n=6)$.
Weekly analytical workload varied from $>20$ hours ($n=6$), $11$--$20$ hours ($n=3$), $5$--$10$ hours ($n=9$), to $<5$ hours ($n=2$). Self-claimed familiarity with immersive technology ranged from \textit{not familiar} $(n=2)$ and \textit{basic} $(n=11)$ to \textit{intermediate} $(n=5)$ and \textit{expert} $(n=2)$. Collaborative software were used \textit{daily} $(n=18)$, \textit{weekly} $(n=1)$, or \textit{occasionally} $(n=1)$ in participants’ work practices. Table~\ref {tab: participant details} summarizes key aspects of participants’ work practices, demonstrating their diversity.

\begin{figure*}[t]
    \centering
    \includegraphics[width=0.8\linewidth]{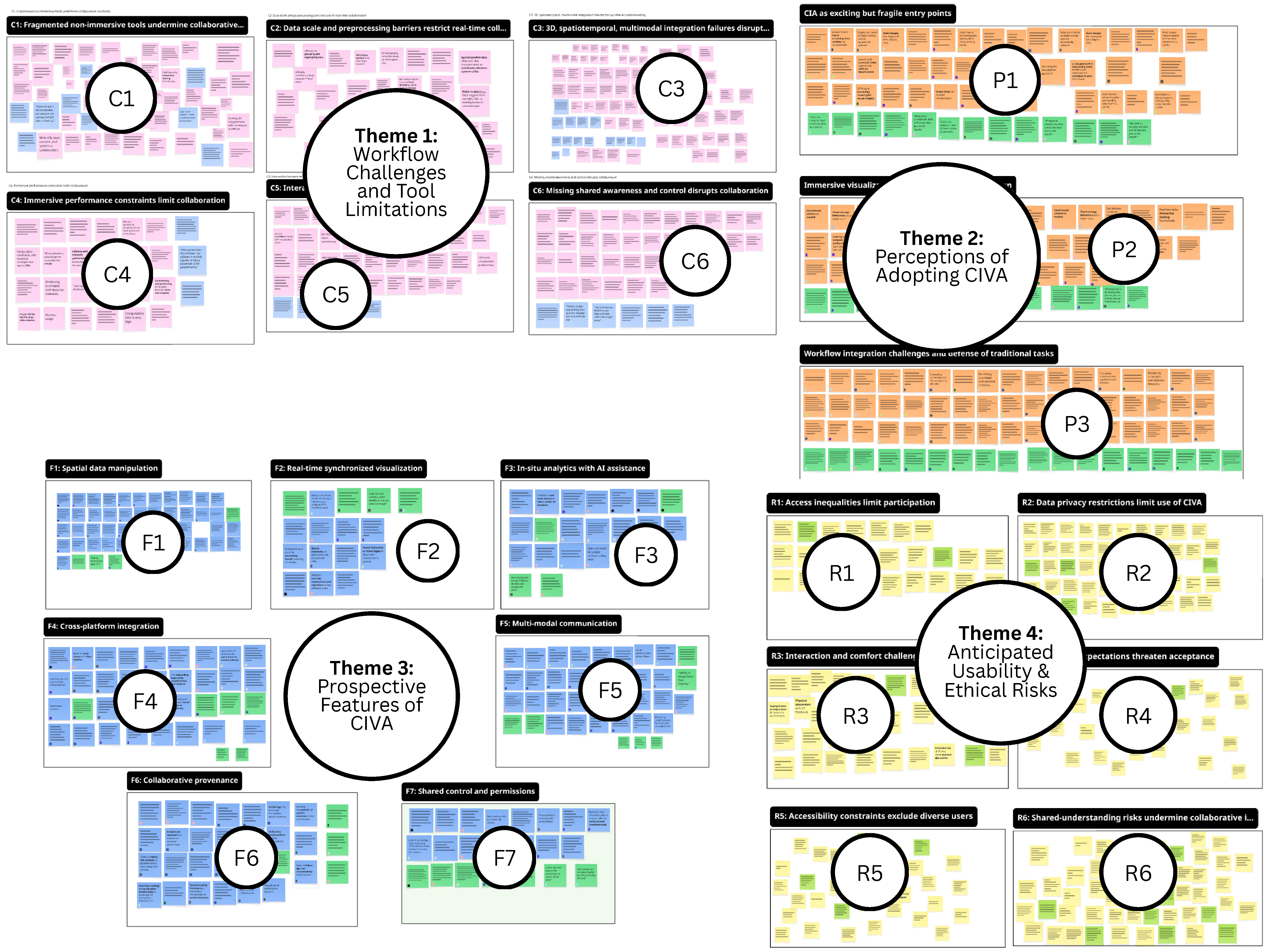} 
    \caption{Our thematic analysis resulted in four major themes, each comprising multiple sub-themes.}
    \Description{This figure illustrates the thematic structure derived from our qualitative analysis. 
    Four major themes were identified, each grouping several sub-themes that capture specific patterns in 
    participant responses. The visualization shows how sub-themes cluster within their respective themes, 
    highlighting overlaps and relationships across categories. The four major themes include: 
    (1) Workflow Challenges and Tool Limitations, covering fragmented practices, steep learning curves, and 
    interoperability issues; 
    (2) Perceptions on Adopting CIVA, capturing expectations, concerns, and conditions influencing adoption 
    across co-located, geo-distributed, and hybrid teams; 
    (3) Prospective Features and Desired Capabilities, such as multi-user annotation, provenance capture, 
    and flexible data integration; and 
    (4) Anticipated Risks and Ethical Concerns, including data privacy, accessibility inequities, and 
    long-term sustainability. 
    Together, these themes provide a structured lens on participants' perspectives and inform design 
    implications for future CIVA systems.}
    \label{fig:themes}
\end{figure*}

%%%%%%%%%%%%%%%%%%%%%%
\subsection{Interview Questions and Procedure}
%%%%%%%%%%%%%%%%%%%%%%
\begin{table}[t]
\centering
\small
\begin{tabular}{p{2.8cm}|p{3.1cm}|l}
\textbf{Video Demo Title} & \textbf{Content Type} & \textbf{Date Posted} \\
\hline
ParaView XR Interface & Scientific Visualization Demo & Nov 30, 2023 \\
\hline
ImAxesGEO & Geoscience Data Analytics Demo & Oct 19, 2022 \\
\hline
Uplift & Collaborative Tabletop \newline Visualization Demo & Oct 20, 2020 \\
\hline
EarthGraph & Seismic / Geoscience \newline Domain Visualization & Sep 26, 2020 \\
\hline
Embodied Axes & AR Interaction Technique Demo & Apr 23, 2020 \\
\hline
FIESTA & Collaborative Immersive Analytics System & Nov 10, 2019 \\
\hline
Data Visualization in VR & Data Cluster Interaction & May 13, 2017 \\
\hline
VR-Assisted Microscopy Data Visualization & Biomedical Microscopy \newline Visualization Demo & May 4, 2016 \\
\hline
HammerheadVR & Genomics Data Exploration Demo & Aug 3, 2015 \\
\hline
See Through Brains & Neuroscience / Brain \newline Visualization Demo & Apr 10, 2013 \\
\end{tabular}
\caption{Overview of video demos categorized by title, domain-specific content type, and posting date (sorted newest to oldest). Links to the corresponding videos are provided in Appendix~\ref{Video_Link}.}
\label{tab:video-demos}
\end{table}

The interview questions (Supplement A) are scripted to address the research questions and organized into four parts that progress from participants’ current practices with high-dimensional data to prospective, collaborative uses of IA. \textbf{Part 1} introduces participants’ roles, domains, and interactions with high-dimensional data, covering routine analysis and visualization practices, toolchains used individually and collaboratively, and recurring breakdowns relevant to collaborative system design. \textbf{Part 2} captures participants’ views on integrating immersive technologies into existing workflows, focusing on integration points, anticipated benefits and burdens, multi-user coordination, and required capabilities. \textbf{Part 3} uses short video demonstrations as design probes: we compile $10$ clips of existing visualization and analytics systems using public sources\footnote{YouTube \href{https://www.youtube.com}{- https://www.youtube.com/}} and conference websites listed in Table~\ref{tab:video-demos} into a single $213$-second video (Supplement B), trimming each by $\approx 12$–$18$ seconds to foreground key interaction and visualization concepts. Participants watched the same compilation once; although they did not directly interact with the technologies shown. Their substantial experience in scientific visualization and analytics, combined with varying levels of familiarity with immersive systems, enabled informed insights about expected use, collaboration patterns, and potential challenges. The compiled video was used only as design probes, with themes reflecting participants’ broader experiences, not evaluations of the interfaces, and the format also supported remote interviews. \textbf{Part 4} concludes with final insights, including physical discomfort, accessibility that may influence adoption, and sustained use. The interview followed a consistent five-step structure, as illustrated in Figure~\ref{fig:interview_stages}.
\begin{itemize}[leftmargin=*]
    \item \textbf{Orientation}     ($\approx$5~min): The facilitator introduced the study, outlined objectives and procedures, and confirmed informed consent.
    \item \textbf{Background \& Practices} ($\approx$10 min): Participants described their roles, the data they work with, the tools and workflows they use, and the recurring challenges they face when analyzing or visualizing high-dimensional data, including collaboration-specific limitations.
    \item \textbf{Perspectives} ($\approx$20 min): Participants discussed expected benefits and risks, task suitability, and required capabilities for effective collaborative analysis.
    \item \textbf{Demo \& Design Probes} ($\approx$20 min): After viewing the video, participants offered additional insights on earlier discussions and reflected on collaborative features, anticipated hurdles and mitigations, and practical considerations.
    \item \textbf{Conclusion }($\approx$5 min): Participants shared final reflections and recommendations for collaborative features. 
\end{itemize}

%%%%%%%%%%%%%%%%%%%%%%
\subsection{Data Analysis Process}
%%%%%%%%%%%%%%%%%%%%%%
We coordinated the analysis remotely in Miro\footnote{Miro \href{https://miro.com/}{- https://miro.com/}}, where all materials and notes were maintained in a shared workspace. Each interview was audio-recorded and transcribed verbatim. We followed a hybrid deductive-inductive thematic analysis approach~\cite{fereday2006demonstrating, xu2020applying}, beginning with a deductive pass in which the lead researcher coded transcripts using a code manual derived from the research questions and interview guide. The research team then met online to compare codes, discuss interpretations, and examine how disciplinary backgrounds shaped analytic judgments. Instead of calculating a statistical inter-rater reliability (IRR) for the analysis, consensus among the team was reached through daily meetings, focused workshops, and discussions~\cite{mcdonald2019reliability}. We then converted the coded data into  2,153 digital statement cards, each containing the ID (P1–P20) and either the original participant quote or a concise, summarized phrase derived from it. With the code manual still too broad, we conducted an inductive coding phase~\cite{braun2012thematic, terry2017thematic} using affinity diagramming~\cite{harboe2015real} to refine broad categories, surface sub-themes, and consolidate the thematic structure, as shown in Figure~\ref{fig:themes}. We ensured trustworthiness (cf.,~\cite{nowell2017thematic}) by involving multiple researchers, documenting analytic decisions, and retaining boundary-case excerpts that refined emerging themes. Supplement C presents the code manual evolution, final themes, and all study memos.

%%%%%%%%%%%%%%%%%%%%%%
\section{Findings}\label{finding}
%%%%%%%%%%%%%%%%%%%%%%
Sections~\ref{Workflow} and~\ref{Collaboration} outline current workflows and collaborative practices, followed by Sections ~\ref{Theme1} (Theme 1), ~\ref{Theme2} (Theme 2), ~\ref{Theme3} (Theme 3), and ~\ref{Theme4} (Theme 4), which present our thematic findings organized into three broad areas for clarity, as shown in Table~\ref{tab:theme_mapping}. 

%\bo{the table is too far away from the text.}

\begin{table}[t]
\centering
%\footnotesize
\resizebox{\columnwidth}{!}{
\begin{tabular}{p{2.9cm}|c|c|c|c}
\textbf{Field} 
& \textbf{T1} 
& \textbf{T2} 
& \textbf{T3} 
& \textbf{T4} 
\\
& \textit{Challenges}
& \textit{Perceptions}
& \textit{Features}
& \textit{Risks}
\\ \hline

\textbf{Tools \& Data \newline Foundations}
& C1, C2 
& P1 
& F2, F3, F4 
& R1, R2 
\\ \hline

\textbf{XR System \& Spatial \newline Capabilities}
& C3, C4 
& P2 
& F1, F2 
& R3, R4 
\\ \hline

\textbf{Human \newline Interaction}
& C5, C6 
& P3 
& F5, F6, F7 
& R5, R6 
\end{tabular}}
\caption{Overview of how each domain area maps to core themes across workflow challenges, perceptions, prospective features, and anticipated risks associated with adopting collaborative immersive visualization and analytics.}
\vspace{-0.5cm}
\label{tab:theme_mapping}
\end{table}

%%%%%%%%%%%%%%%%%%%%%%
\subsection{Current Visualization \& Analytics Ecosystem} \label{Workflow}
%%%%%%%%%%%%%%%%%%%%%%

\begin{figure}[t]
    \centering
    \includegraphics[width=0.86\linewidth]{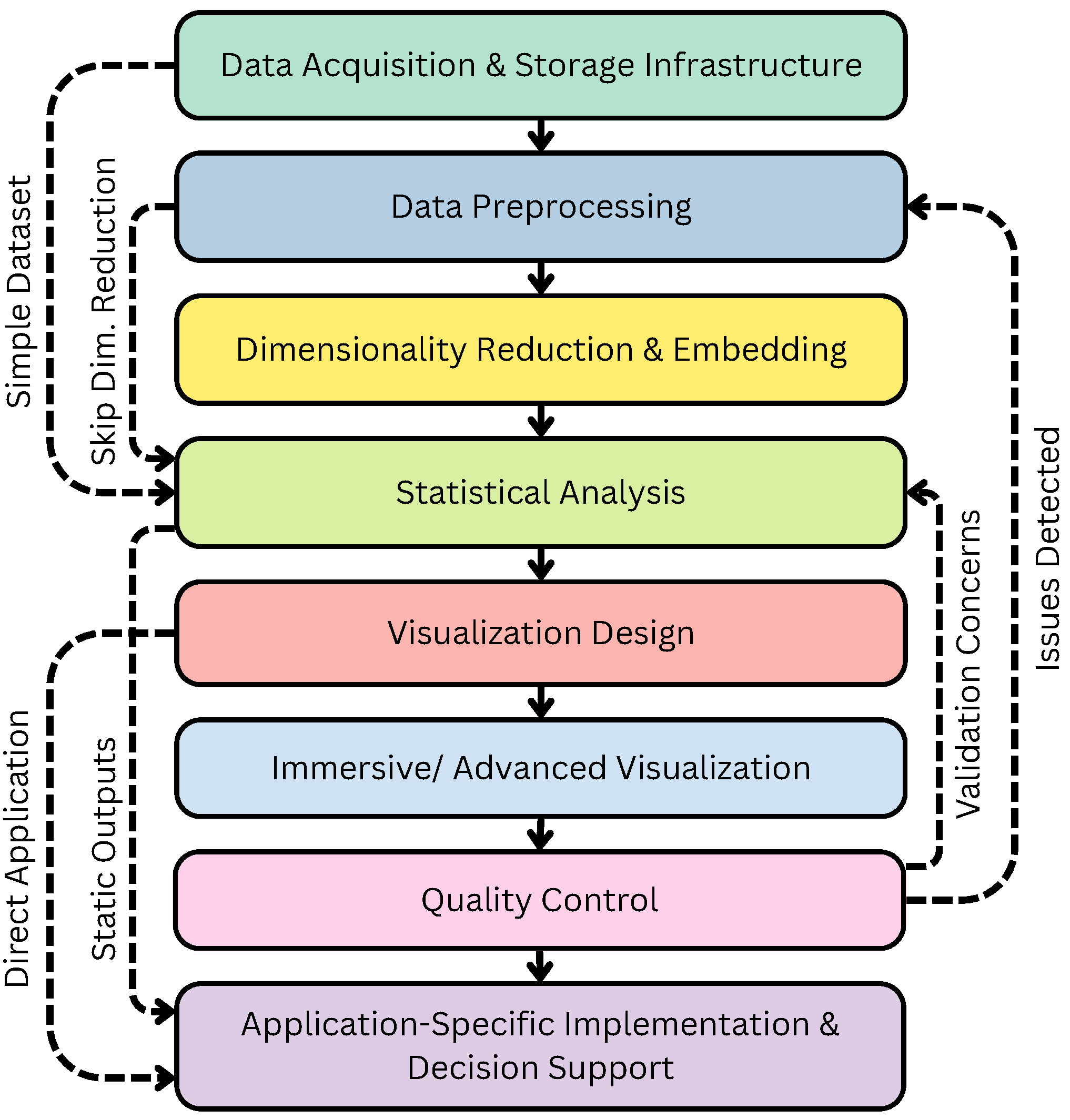}
    \caption{Current workflow for high-dimensional scientific data visualization and analytics.}
    \Description{The workflow begins with data acquisition and storage infrastructure, ensuring integration of heterogeneous datasets. 
    Data preprocessing then prepares inputs for analysis, followed by dimensionality reduction and embedding techniques to simplify high-dimensional data while preserving structure. 
    Statistical analysis enables hypothesis testing and model development, which inform visualization design choices. 
    Immersive and advanced visualization approaches (e.g., VR and AR) allow domain experts to interact with complex datasets. 
    Quality control validates outputs, ensuring reproducibility and accuracy. 
    Application-specific implementation and decision support translate insights into actionable knowledge for scientific or policy contexts. 
    Feedback loops emphasize the iterative nature of the workflow: issues identified in later stages may require revisiting earlier steps, 
    while simpler datasets can sometimes bypass intermediate processes for direct application.}
    \label{fig:workflow}
\end{figure}

Synthesized from interview findings, Figure~\ref{fig:workflow} illustrates the workflow practices of domain experts in visualizing and analyzing high-dimensional scientific data, and Table~\ref{tab: participant details} summarizes their primary analysis and visualization tools. 
\footnote{Technical tools, frameworks, software, and methods mentioned in this section are listed in the supplementary document.}
%\bo{the table is too far away from the text.} \fan{This table has been reused. First mentioned in section 3.2, that's why it is there.} 

\subsubsection{\textbf{Data acquisition \& storage infrastructure.}} Domain experts report acquiring data from heterogeneous sources, ranging from structured databases and unstructured streams to satellite imagery, LiDAR scans, camera-trap photos, audio recordings, scientific waveforms, medical imaging systems, and real-time HPC telemetry. Several note infrastructures, where REDCap and PACS manage clinical and imaging data across numerous projects. Storage practices span local HDDs or SSDs for initial storage, cloud services like OneDrive and Google Drive for backup, and HPC clusters or AWS with Kubernetes for large-scale processing, often supported by distributed frameworks such as Dask. Experts require navigating frequent format transitions—moving from CSV for quick inspection to Parquet or Arrow for efficient processing, or to databases like DocDB for high-dimensional search, and one participant remarks that \quoting[P11]{Every format switch brings metadata inconsistencies and tool-specific constraints}.

\subsubsection{\textbf{Data preprocessing}} is identified by participants as a critical transformation stage that cleans and restructures high-dimensional data before analysis.
Several noted that early wrangling can be demanding, explaining that \quoting[P6]{it's difficult if the data has some missing values}. Experts rely on Python frameworks like Pandas, NumPy, and Dask for distributed processing, and high-performance DataFrame libraries such as Polars and Arrow—which \quoting[P10]{allow us to efficiently move between data processing engines without copying memory}, alongside VTK-based workflows for LiDAR and other sensor data. Traditional tools such as Excel, Stata, and R remain common for initial cleaning and statistical preparation.

\subsubsection{\textbf{Dimensionality reduction \& embedding}} is described by participants 
as essential steps for making dense, multi-dimensional datasets workable and interpretable, even though \quoting[P13]{...lose some of the semantics of the data}. Experts routinely reduce high-dimensional vectors using t-SNE or UMAP, with one noting they \quoting[P10]{moved to UMAP because it better captured the patterns}. Participants also generate latent representations using autoencoders and produce cross-modal embeddings through neural networks and large language models. They rely on TensorFlow for model training and R-based Linux clusters for scalable computation, while custom visualization tools, such as quaternion unit-sphere displays and D3.js frameworks, are used to compute and interpret reduced representations.

\subsubsection{\textbf{Statistical analysis}} is the stage where participants report making cleaned data analytically meaningful through statistical modeling, machine learning, and AI-assisted inference. Many rely on time-series techniques, with one mentioning the use of \quoting[P12]{matrix profiles and ARIMA analysis to make predictions and identify anomalies}. Others apply AI-driven classification or segmentation to extract structure from imaging, audio, or ecological data, and several employ regression or machine-learning models after preprocessing. Their computational toolkit spans Python and TensorFlow for neural networks, R and RStudio for visualization and descriptive statistics, MATLAB for scripting and plotting, and Stata for large statistical analyses. Human oversight remains essential, as participants emphasized that visualization must \quoting[P16]{...enable the user to interact and correct} AI-derived results.

\subsubsection{\textbf{Visualization design}} appears in participants’ workflows as the step where high-dimensional analyses are translated into interpretable visual structures using scientific, statistical, and web-based tools.
Several rely on ParaView, noting \quoting[P20]{...easier to load mesh data and apply filters like smoothing or point-flagging}, when working with 3D, LiDAR, or hierarchical structures. Python tools such as Matplotlib, Altair, and Vega-Lite support plots, heatmaps, and correlation matrices, while high-performance workflows pair with Polars for large DataFrame operations. Domain-specific representations include unit-sphere views for rotational motion, velocity maps, and 2D heatmaps, as noted, \quoting[P8]{Still, it’s really hard to show 3D visualizations}. Interactive displays often use D3.js and Observable for richer customization, and Stata, R, MATLAB, Tableau, and Plotly provide publication-ready figures and stakeholder dashboards for reporting.

\subsubsection{\textbf{Immersive \& advanced visualization}} emerges in participants’ workflows as the stage where complex data is explored within spatial, interactive environments designed to \quoting[P6]{enable the analysis of the data in a more effective way, better than just the monitor}. Teams used Unity or Unreal Engine to build VR digital twins, and several experimented with NVIDIA Omniverse for collaborative 3D development, alongside specialized tools such as LiDARViewer with the Rui VR toolkit for scientific inspection. Their infrastructure often utilizes Omniverse Nucleus to keep 3D scene states synchronized across collaborators.
Integration pipelines connected QGIS and Python to Unreal Engine for API-driven updates, allowing experts to \quoting[P18]{change basic features like color or its material} or modify text dynamically. Custom plugins enable CAVE immersive displays, while WebXR and commercial tools support browser-based immersion.

\subsubsection{\textbf{Quality control}} surfaces in participants’ practices as the process of ensuring computed results align with raw data and expected outcomes. Experts emphasized contextualizing outputs against raw inputs, noting the need to \quoting[P16]{show the computational results in the context of the raw data for QC}. When multimodal data lack precise registration, participants rely on manual inspection, explaining that \quoting[P15]{most people just do side-by-side views and let the human eye take care of alignment}. QC routines include reproducibility checks by external and collaborative inspection through TensorBoard. Supporting infrastructure spans hierarchical data models, standardized web portals, and pseudonymized medical workflows, with domain-specific tasks using annotation tools to verify whether outputs match expected results.

\subsubsection{\textbf{Application-specific implementation}} appears in participants’ accounts as the stage where analytical results integrate into domain workflows, ranging from app monetization to sports analytics. Disaster researchers combine \quoting[P19]{GIS with QGIS and Python and translate its visuals in Unreal Engine}, while digital-twin teams map supercomputer telemetry into immersive environments using Omniverse to \quoting[P18]{generalize data-center behavior from the streams coming off the Frontier system}. Supporting infrastructures include Tableau dashboards, custom D3 and Observable systems for scientific interactivity, Quarto for live HTML reports, and pipelines that merge QGIS outputs with markdown narratives across environmental, healthcare, and policy applications.

%%%%%%%%%%%%%%%%%%%%%%
\subsection{Current Collaboration Infrastructure} \label{Collaboration}
%%%%%%%%%%%%%%%%%%%%%%

\subsubsection{\textbf{Team communication}} relies on platforms such as Zoom or Microsoft Teams for synchronous discussion and Slack or email for asynchronous coordination, yet these remain disconnected from analytical environments. Participants struggled to convey insights about high-dimensional data through text-based channels, noting that \quoting[P17]{When sharing static screenshots, it’s very hard to explain my perspective - they can’t interact with the data}. This disconnect fragments workflows and causes critical context to be lost. Another participant explained, \quoting[P12]{Most platforms don’t let us talk over the same visualization in real time, so we end up describing things in words, which creates confusion}. Experts are forced to translate complex visual patterns into verbal descriptions, increasing misinterpretation and prolonging collaboration.

\subsubsection{\textbf{Sharing \& distribution methods}} predominantly involve exporting static images via email or cloud storage, removing the interactivity vital for scientific analysis. As one participant noted, \quoting[P19]{I just export a PNG or PDF, but then the interactive aspect is gone, so my collaborators can’t explore the data themselves}. File-based sharing through Google Drive or Dropbox adds further friction due to limited integration with domain-specific visualization tools. The asynchronous nature
exacerbates the problem, with another expert explaining, \quoting[P11]{We send Excel sheets or CSVs over email, and by the time someone plots them, it’s already outdated compared to my version}. Such practices force repeated format conversions, contributing to inefficiencies and data integrity issues. Some teams use Jupyter or Google Colab to rerun code, but they still lack synchronous manipulation.
 
\subsubsection{\textbf{Version control}} currently relies on manual file naming and direct communication, offering no systematic means to track changes or maintain provenance. As one participant explained, \quoting[P16]{If two of us modify the same dataset, it’s impossible to track which visualization is the latest without asking directly}. While some experiment with GitHub, its file-size limits and the scalability constraints of Git LFS  make it unsuitable for high-dimensional datasets, pushing teams toward ad-hoc workarounds such as appending version numbers, an approach one expert described as \quoting[P18]{version control becomes nightmare with multiple files (v1, v2, v3)}. Some teams use shared HPC servers to edit the same script and verify results by rerunning analyses across environments, \quoting[P11]{We share our code with them, and they run the code in their environment}, highlighting the inconsistency of provenance tracking.

%%%%%%%%%%%%%%%%%%%%%%
\subsection{Theme 1: Workflow Challenges and Tool Limitations}\label{Theme1}
%%%%%%%%%%%%%%%%%%%%%%

Having outlined current workflows and practices, we now examine the challenges experts encounter in their everyday work with high-dimensional scientific data.

\subsubsection{\textbf{C1:}} \textbf{Fragmented non-immersive tools undermine collaborative continuity} by limiting teams’ ability to jointly manipulate high-dimensional visualizations in real time. Participants emphasized the absence of a shared workspace, noting that current tools are \quoting[P1]{split into different areas and different purposes}, forcing collaborators into parallel, disconnected flows. Lacking real-time interactivity, teams resort to static exchanges where \quoting[P17]{sharing static images (PNGs) hinders dynamic discussion and real-time interaction}. Even synchronous meetings remain constrained, as tools like ParaView require exporting snapshots, leading one researcher to repeatedly \quoting[P2]{save, convert, and send files} instead of co-exploring a common view. Parallel toolchains across Python, R, and ParaView further complicate collaboration, making it difficult to maintain visual coherence and shared understanding.

\subsubsection{\textbf{C2:}} \textbf{Data scale and preprocessing barriers restrict real-time collaboration} because high-dimensional datasets often exceed the capacity of the tools that collaborators rely on. Participants described struggling to even render data before joint work begins, noting that \quoting[P17]{the first challenge is getting it to actually show up} and that terabyte-scale inputs routinely overwhelm cluster interfaces or desktop tools. Such limits prevent synchronous exploration, since \quoting[P2]{we need to subsample and scale it down to the point that the visualization system can even run}. Immersive platforms intensify these concerns, with researchers fearing that VR would \quoting[P4]{cause even a higher computational load} when 2D tools already fail on large files. As a result, preprocessing steps (conversion, reduction, and formatting) must occur outside collaborative sessions, because \quoting[P7]{the software for data visualization and immersive visualization is not streamlined}, breaking the continuity needed for joint sensemaking.

\subsubsection{\textbf{C3:}} \textbf{3D, spatiotemporal, and multimodal integration failures disrupt shared understanding} by preventing collaborators from maintaining a shared view of complex scientific data. Teams repeatedly described how 3D structures collapse into flat, non-interactive representations, making it impossible to communicate perspective, as noted, they \quoting[P2]{cannot show the data from a particular angle unless I take screenshots}. Even fundamental spatial reasoning breaks down because, as another explained, \quoting[P4]{we try to visualize 3D data on a 2D screen, which is a challenge}. Multimodal data introduces additional misalignment: analysts struggle to correlate telemetry, video, and derived parameters, with one mentioning \quoting[P8]{it's difficult to combine information from cameras and other tools to tell the same story}.

\subsubsection{\textbf{C4:}} \textbf{Immersive performance constraints limit collaboration} because high-dimensional datasets exceed what headsets and real-time renderers can support, compounded by setup time, heterogeneous hardware, and inconsistent performance. Participants already experienced overload in 2D workflows, noting that \quoting[P4]{I'm already having trouble loading large files, so VR will cause even a higher computational load}. Immersive systems further amplify these bottlenecks when computation shifts onto the device; as one expert mentioned, \quoting[P5]{the computation resource is limited, so there’s a tradeoff}. Consequently, experts reported downsampling or converting data before collaborative sessions because \quoting[P6]{current pipelines and devices cannot handle full-resolution inputs}.

\subsubsection{\textbf{C5:}} \textbf{Interaction barriers reduce collaborative precision} by limiting the accuracy, expressiveness, and endurance required for joint analysis. Participants noted that gesture and controller-based input lacks the fine control needed for scientific tasks, explaining that \quoting[P1]{very hard to interact with because we can only use our fingers}. Such imprecision made collaborators uncertain about each other’s selections or actions in shared XR spaces. Entering numerical values or equations further strained workflows, as users found it \quoting[P14]{pretty hard for us to write some text,...to write equations}. Physical discomfort further restricts collaborative depth, as prolonged immersion leads to symptoms that \quoting[P4]{will definitely hamper my problem-solving skills and critical thinking}.

\subsubsection{\textbf{C6:}} \textbf{Missing shared awareness and control disrupts collaboration} by preventing teams from establishing a stable, shared frame of reference in immersive environments. Participants noted that collaborators cannot easily confirm mutual alignment, wondering \quoting[P13]{...whether they see the same thing as I do}. Without shared indicators or synchronized viewpoints, collaborators must resort to verbal directions, \quoting[P2]{I need my colleague to see the data from a particular angle, unless I tell them}. Control conflicts further impede coordination, and the absence of persistent records limits continuity, with participants noting that they \quoting[P4]{can't see the history of what other colleagues are doing}.

\vspace{0.2cm}
\camera{These challenges reflect the perspectives of experienced domain experts working in distributed, interdisciplinary teams, where misalignment can directly affect analytical work. Participants often described collaboration in terms of keeping everyone aligned across people, tools, and ongoing analysis, reflecting shared responsibility rather than individual task ownership. This emphasis was especially pronounced in accounts of geographically 
distributed work, where maintaining coordination requires explicit effort. The interview-based format, which relied on reflection rather than direct observation, likely brought these coordination concerns to the foreground compared to more moment-to-moment interaction issues.}

%%%%%%%%%%%%%%%%%%%%%%
\subsection{Theme 2: Perceptions of Adopting CIVA}\label{Theme2}
%%%%%%%%%%%%%%%%%%%%%%

We next examine how domain experts perceive the adoption of CIVA, capturing their initial impressions, expectations, and concerns about its role in their work.

\subsubsection{\textbf{P1:}} \textbf{CIVA as exciting but fragile entry points} reflects participants’ mixed first impressions, pairing enthusiasm with concerns about practical viability. Many saw clear promise, describing CIVA as \quoting[P19]{very exciting, something required and important to have} and noting that moving within data \quoting[P17]{is going to make it very easy to share information}. Yet others questioned its added value, suggesting \quoting [P3]{little additional value in these interactive animations} for task-oriented work. Practical friction quickly tempered excitement as participants stressed that \quoting[P16]{if the setup takes longer than 10 seconds, nobody in our domain will use it}. For some, the technology still felt \quoting[P1]{very interesting, like a toy}, revealing a gap between CIVA’s perceived potential and the fragility of its integration into established workflows.

\subsubsection{\textbf{P2:}} \textbf{Immersive visualization as a catalyst for collaboration} captures participants’ belief that shared immersive space can strengthen collective reasoning. Immersion was described as creating \quoting[P1]{It will feel like the same workspace, but with the immersive sense of being together}, enabling collaborators to \quoting[P10]{better brainstorm to develop the intuitions together} and build \quoting[P6]{a real deep understanding of the problem}. Participants emphasized immersive environments support communication across levels of expertise, allowing scientists to present insights \quoting[P11]{much more beneficial to the end users and stakeholders} who may not follow traditional plots. Multi-user interaction was seen as central, where \quoting[P5]{Many people can interact with the data, annotate the region of interest, [and] contrast and compare between different selections}. Several envisioned low-friction collaboration as \quoting[P13]{a Google doc for 3D data}, where remote and co-located participants contribute within a shared spatial workspace in real time without moving files across platforms.
 
\subsubsection{\textbf{P3:}} \textbf{Workflow integration challenges and defense of traditional tasks} reflect participants' doubts about the productivity value of immersive environments while defending established workflows. Several raised skepticism about productivity gains, by cautioning \quoting[P10]{[IA] could be counterproductive because it's so immersive and so cool that people will get distracted}. Precision-demanding tasks were described as poorly suited to XR, where interactions are \quoting[P1]{very hard to interact with the data if we can only use our fingers} and raw data processing remains \quoting[P2]{...very hard to process raw data actually}. Participants drew firm boundaries around established practices, emphasizing that \quoting[P18]{programming and scripting is easiest on my laptop} and that preprocessing should remain automated through familiar pipelines—\quoting[P17]{I would prefer just to write a Python script and get them all changed}. These concerns underscore that CIVA must align with, rather than replace, entrenched computational workflows to avoid disrupting productivity.

\vspace{0.2cm}
\camera{These perceptions reflect participants' experience as established practitioners evaluating unfamiliar technology. The tension between enthusiasm and skepticism was shaped by prior encounters with adoption barriers in research settings, particularly among those with well-developed analytical routines. Different user groups, such as novices without entrenched workflows, might respond quite 
differently to the same collaborative affordances.}

%%%%%%%%%%%%%%%%%%%%%%
\subsection{Theme 3: Prospective Features of CIVA}\label{Theme3}
%%%%%%%%%%%%%%%%%%%%%%

\noindent \textbf{Core functionalities} (\ref{F1},~\ref{F2},~\ref{F3},~\ref{F4}) are foundational for enabling effective collaboration, allowing immersive systems to overcome 2D limitations and support high-dimensional data analysis. \camera{Interviewees emphasized that these capabilities must be in place before collaborative mechanisms can function reliably, as shared understanding and coordinated action depend on stable, responsive analytical foundations.}

\subsubsection{\textbf{F1:}} \textbf{Spatial data manipulation}\label{F1} enables intuitive exploration by letting collaborators interact with high-dimensional data through multi-modal gestures and coordinated actions. Participants stressed the value of hands-on control, noting that users can \quoting[P1]{use our fingers or the controllers to interact with data} for operations such as lassoing, zooming, filtering, and adjusting dimensions. Spatial interaction also streamlines collaborative data cleaning, allowing teams to \quoting[P1]{directly remove that data} during early analysis. Effortless manipulation is critical for maintaining focus, with participants calling for interactions that \quoting[P15]{help you manipulate the data in a very smooth and easy way} without interrupting scientific reasoning.

\subsubsection{\textbf{F2:}} \textbf{Real-time synchronized visualization}\label{F2} keeps teams aligned by continuously updating shared scenes as the underlying data evolves. Participants stressed the need for high-frequency streaming, where \quoting[P16]{there is no break in updating the visualizations, as everything has to be streaming-based}, ensuring collaborators remain synchronized with ongoing scientific measurements. Temporal controls further enhance collaborative reasoning by letting teams inspect evolving signals, \quoting[P17]{show the time variance of the waveform, not just statics in image} and revisit prior states through replay features that allow teams to \quoting[P5]{record all your steps while you do the data analysis, and replay the whole process} for validation or comparison. 

\subsubsection{\textbf{F3:}} \textbf{In-situ analytics with AI assistance}\label{F3} provides dynamic in-situ analytics that brings computation into immersive space by enabling users to initiate transformations, filtering, and statistical operations directly inside CIVA. Participants highlighted the value of immediate computation, \quoting[P17]{press button and run like T-SNE on the data and visualize the result} which eliminates context switching and speeds decision-making during collaborative sessions. AI support enhances this workflow by translating high-level scientific intent into concrete operations, allowing assistants to \quoting[P12]{produce low-level actions that can be useful for updating… the visuals}. Together, these capabilities position CIVA as an active analytical workspace.

\subsubsection{\textbf{F4:}} \textbf{Cross-platform integration}\label{F4} preserves existing workflows by embedding immersive visualization within established scientific tools rather than replacing them. Participants stressed that CIVA must fit naturally into environments built around Python and Jupyter, envisioning \quoting[P12]{a library in Python that allows you to create immersive visualizations… instead of using Plotly or matplotlib}. Seamless movement between desktop and immersive contexts supports hybrid work patterns and allows collaborators to mix immersive exploration with conventional scripting, while also providing \quoting[P16]{continuity between the realities} needed to avoid fragmenting multi-stage scientific analysis.

\vspace{0.2cm}
\noindent\textbf{Collaborative features} (\ref{F5},~\ref{F6},~\ref{F7}) play the central role in supporting shared understanding, allowing immersive systems to provide synchronized interaction and effective teamwork in scientific data analysis and visualization.

\subsubsection{\textbf{F5:}} \textbf{Multi-modal communication}\label{F5} sustains shared understanding by combining voice, gestures, avatars, and visual indicators to maintain common focus in immersive sessions. Voice remained central because \quoting[P8]{voice is the most efficient way for real-time exchange}, while gestures helped synchronize attention by letting collaborators \quoting[P15]{show colleagues what to look at}. Participants emphasized the need for telepointers and focus indicators, since simple verbal cues like “look here” fail to identify precise features; instead, each collaborator should have distinguishable markers such as colored lines or labeled cursors. Avatars further support mutual awareness, helping people \quoting[P1]{feel like the same workspace} and reinforcing presence in shared spatial reasoning.

\subsubsection{\textbf{F6:}} \textbf{Collaborative provenance}\label{F6} preserves shared memory by capturing actions, annotations, and reasoning steps across sessions. Participants stressed the need to \quoting[P4]{make notes or drop a breadcrumb, knowing that I’m not going to lose it}, ensuring insights remain accessible for future collaborators. Provenance requires linking actions to individuals so teams can \quoting[P11]{know that this was something you did or something I did}, improving accountability and traceability during group exploration. Rich session history—including transcripts of discussions and the ability to \quoting[P12]{redo or undo their actions and move back and forth in time} supports replication, review, and onboarding without repeating prior analyses.

\subsubsection{\textbf{F7:}} \textbf{Shared control and permissions}\label{F7} coordinate multi-user interaction by regulating simultaneous manipulation, defining roles, and ensuring equitable participation across devices. Participants warned that without structured control, immersive collaboration becomes chaotic, noting the need to \quoting[P3]{only one person can move it when you are presenting the data} to prevent accidental interference. Permission models further safeguard sensitive work by allowing teams to \quoting[P1]{not give all the permissions to access all the data}, aligning access with responsibility and expertise. Cross-device compatibility ensures 2D collaborators remain included, addressing concerns that \quoting[P10]{if someone uses the devices and some collaborators use 2D devices, it’s less valuable}, reinforcing equity in distributed scientific teams.

\vspace{0.2cm}
\camera{Participants’ proposals for future CIVA systems reflect their experience working as expert analysts in diverse teams. They described features mainly as ways to help people stay aligned while working together, not as tools to replace individual analysis. Different groups, such as novice users or teams working in the same physical space, might value other kinds of support under similar system designs. Overall, these findings reflect how collaboration is understood in expert scientific practice, where immersive environments are positioned as infrastructure for collective sensemaking rather than extensions of individual workspaces.}

\begin{table*}[t]
\centering
\setlength{\tabcolsep}{5pt}
\renewcommand{\arraystretch}{1.15}
\begin{tabular}{c|c|p{0.25\textwidth}|p{0.25\textwidth}|p{0.25\textwidth}}
% \hline
\multicolumn{2}{c|}{} & \multicolumn{3}{c}{\textbf{Place}} \\
\cline{3-5}
\multicolumn{2}{c|}{} & \textbf{Same Place} & \textbf{Different Place} & \textbf{Hybrid} \\
\hline
\multirow{2}{*}{\textbf{Time}}
& \textbf{Same Time}
& \ref{I1} Coordinating Work through Shared Computational History \vspace{0.2cm} \newline
  \ref{I2} Enhancing Awareness through \newline Social Translucence
& \ref{I3} Synchronizing Cross-Device \newline Workflows for Hybrid Presence \vspace{0.2cm} \newline
  \ref{I5} Orchestrating Collaboration through AI-Mediated Coordination
& \ref{I3} Synchronizing Cross-Device \newline Workflows for Hybrid Presence
\\ \cline{2-5}
& \textbf{Different Time}
& \ref{I1} Coordinating Work through Shared Computational History
& \ref{I4} Integrating Accessibility into \newline Collaborative Infrastructures
& \ref{I4} Integrating Accessibility into \newline Collaborative Infrastructures
\\
% \hline
\end{tabular}
\vspace{0.15cm}
\caption{Extended CSCW Time–Space Matrix showing how our five design implications (\ref{I1},~\ref{I2},~\ref{I3},~\ref{I4},~\ref{I5}) align with synchronous/asynchronous and co-located/distributed/hybrid settings.}
\label{tab:cscw_matrix}
\end{table*}

%%%%%%%%%%%%%%%%%%%%%%
\subsection{Theme 4: Anticipated Usability \& Ethical Risks}\label{Theme4}
%%%%%%%%%%%%%%%%%%%%%%
Beyond prospective features, participants raised concerns about anticipated usability challenges and ethical risks that could shape the adoption of CIVA.

\subsubsection{\textbf{R1:}} \textbf{Access inequalities limit participation} because participants expect uneven availability of XR hardware, high device costs, and connectivity constraints to prevent many collaborators from joining CIVA sessions. Several foresee reluctance to adopt XR, noting that \quoting[P10]{people do not have the hardware and… are reluctant to invest in one}, while others anticipate bandwidth issues where \quoting[P11]{Internet bandwidth can just make things laggy}. Such disparities raise concerns about equitable involvement across institutions and regions.

\subsubsection{\textbf{R2:}} \textbf{Data privacy restrictions limit use of CIVA} because participants expect that legal and institutional rules will restrict how sensitive data enters immersive platforms. In medical domains, data cannot leave local infrastructure, since \quoting[P16]{we are not by law allowed to upload the data somewhere}. Participants also anticipate barriers when collaborating across organizations with incompatible policies, such as national laboratories, government agencies, or private companies, as noted, \quoting[P15]{data is useless if we have to share it through third-party apps}.

\subsubsection{\textbf{R3:}} \textbf{Interaction and comfort challenges reduce usability} because participants expect imprecise gesture-based interaction, text-entry difficulty, and ergonomic strain to limit effective analysis. They predict discomfort, noting that they \quoting[P9]{can't be productive with eyestrain and headache}, and foresee that immersive manipulation may feel \quoting[P1]{very hard to interact with if we can only use our fingers}. Such usability burdens raise concerns about sustained analytical work in CIVA.

\subsubsection{\textbf{R4:}} \textbf{Reliability expectations threaten acceptance} because participants expect XR systems to require rapid startup and stable performance to be usable in practice. They caution that adoption will falter if CIVA feels slow or cumbersome, with one stating \quoting[P16]{if the setup takes longer than 10 seconds, then nobody will use it}. These expectations reflect perceived reliability thresholds that CIVA must meet to gain trust.

\subsubsection{\textbf{R5:}} \textbf{Accessibility constraints exclude diverse users} because participants expect ergonomic discomfort, vision-related challenges, and perceptual differences to hinder inclusive adoption. Glasses wearers worry that \quoting[P3]{having a glass on top of your glass, could not work}, while others highlight the needs of collaborators with sensory or motor impairments, stressing that systems must support \quoting[P19]{people who cannot talk or cannot listen or cannot see}. These concerns position accessibility as a core adoption barrier in collaborative settings.

\subsubsection{\textbf{R6:}} \textbf{Shared-understanding risks undermine collaborative interpretation} because participants expect difficulty verifying whether collaborators perceive the same features or viewpoints in immersive space. They worry that it may be \quoting[P10]{difficult to share findings… instead of having a ground truth}, especially when collaborators cannot confirm \quoting[P13]{whether they see and perceive the same thing as I do}. These concerns highlight interpretability as an adoption risk.

\vspace{0.2cm}
\camera{The risks participants anticipated reflect their institutional and disciplinary contexts. Privacy concerns were more salient for those in regulated medical or government settings, while access and reliability issues emerged primarily from distributed teams and production environments. These variations suggest that adoption barriers depend on organizational context, with different settings presenting distinct challenges for CIVA deployment.}

\begin{center} 
\fbox{ 
\parbox{\linewidth}{ 
\noindent To summarize, RQ1 is addressed in Sections ~\ref{Workflow}, ~\ref{Collaboration},~\ref{Theme1}, which describe current workflow and collaboration practices for working with high-dimensional data and the limitations practitioners encounter; RQ2 is addressed in Sections \ref{Theme2} and \ref{Theme4}, which state how experts interpret and anticipate the adoption of CIVA; and RQ3 is addressed in Section \ref{Theme3}, which outlines the features and capabilities participants envision for future systems.} } \end{center}

%%%%%%%%%%%%%%%%%%%%%%
\section{Implications} \label{implication}
%%%%%%%%%%%%%%%%%%%%%%
\begin{table*}[t]
\centering
\begin{tabular}{l | l |  l | l | l}
\textbf{Design Implications} & \textbf{Challenges} & \textbf{Perceptions} & \textbf{Features} & \textbf{Risks} \\

\midrule

\textbf{\ref{I1} Coordinating Work through Shared Computational History} 
& C1, C2, C6
& P1
& F5, F6
& R2, R6 \\

\midrule
\textbf{\ref{I2} Enhancing Awareness through Social Translucence}
& C3, C6
& P2
& F2, F5, F6, F7
& R5, R6 \\
\midrule
\textbf{\ref{I3} Synchronizing Cross-Device Workflows for Hybrid Presence}
& C3, C5, C6
& P3
& F2, F3, F4, F6
& R2, R4, R6 \\
\midrule
\textbf{\ref{I4} Integrating Accessibility into Collaborative Infrastructures }
& C5, C6
& P2
& F1, F4, F5
& R3, R5 \\
\midrule
\textbf{\ref{I5} Orchestrating Collaboration through AI-Mediated Coordination}
& C1, C3, C6
& P1
& F1, F3, F7
& R1, R4, R6 \\
\end{tabular}
\vspace{0.1cm}
\caption{Mapping of proposed design implications to corresponding workflow challenges, prospective features, and anticipated risks associated with adopting CIVA.}
\label{tab:implications_mapping}
\end{table*}

\begin{figure*}[t]
    \centering
    \includegraphics[width=\linewidth]{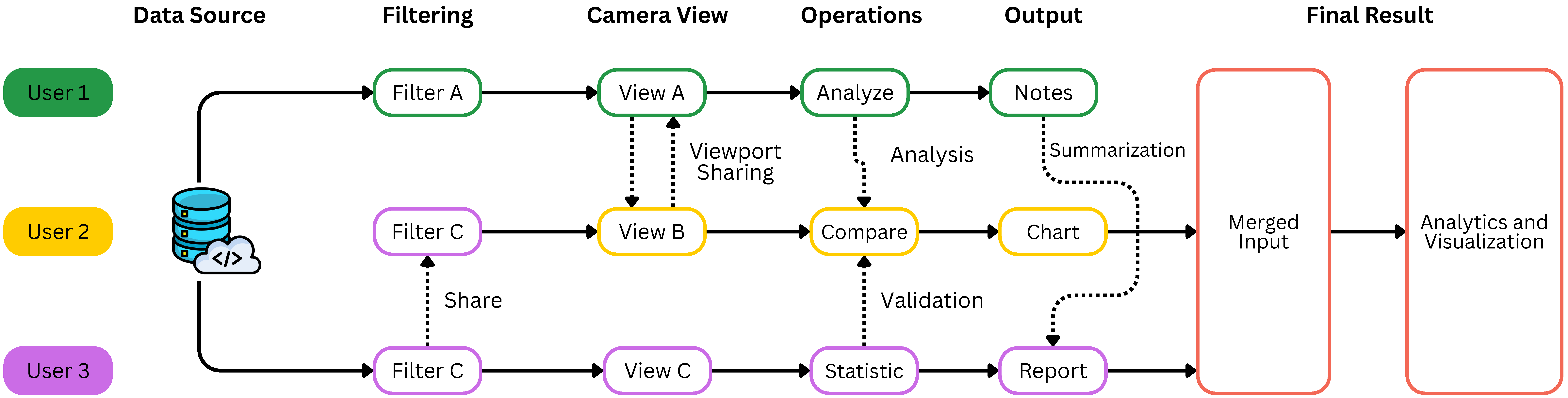}
    \caption{ Provenance swimlanes illustrating a hypothetical collaboration scenario in CIVA, showing branching and merging among three users. Horizontal arrows represent each user’s own workflow progression, while dotted arrows indicate collaborative exchanges. The figure highlights how individual analyses converge into a merged outcome through prospective multi-user provenance tracking.}
    \Description{The provenance swimlanes diagram depicts three parallel lanes for User 1, User 2, and User 3, each progressing from raw data through filtering, visualization, and analysis steps. User 1 contributes annotations and hypotheses, User 2 produces charts and insights, and User 3 performs statistical analysis and reporting. Solid arrows indicate self-interactions within a user’s workflow, while dotted arrows represent collaborative interactions across users, including sharing filters, aligning views, and passing results downstream. These exchanges produce branching and feedback loops, ultimately merging into a unified analysis node on the right. The provenance swimlanes illustrate how provenance captures both individual contributions and cross-user dependencies, ensuring accountability, reproducibility, and transparency in collaborative data analysis.}
    \label{fig:provenance_swimlanes}
\end{figure*}

The observations we identify in CIVA for high-dimensional scientific data reflect foundational issues long examined in Computer-Supported Cooperative Work (CSCW)~\cite{palmer1994computer,grudin2002computer}: coordinating \textit{articulation work}~\cite{schmidt1992taking}, achieving \textit{mutual intelligibility}~\cite{suchman1987plans} in distributed cognitive systems~\cite{hollan2000distributed}, and preserving \textit{interactional coherence}~\cite{heath1992collaboration} across heterogeneous sociotechnical settings. Building on Hutchins’ distributed cognition framework~\cite{hutchins1995cognition}, which views cognition as emerging through interactions among people, artifacts, and environments, these challenges explain why immersive visualization and analytics systems that distribute collaborative reasoning across scientists and tools often falter. While CSCW spans broader themes such as awareness~\cite{dourish1992awareness}, boundary objects~\cite{star1989institutional}, and infrastructures~\cite{bowker2000sorting}, 
our study foregrounds breakdowns in articulation, interpretability, and coherence as the most consequential for CIVA adoption. Recent CSCW research on hybrid scientific collaboration~\cite{neumayr2022territoriality,meyer2025better,bjorn2024achieving} further underscores the persistence of these challenges in immersive analytics contexts~\cite{ens2022immersive}.

\camera{Our findings suggest that collaboration in CIVA cannot be fully explained by frameworks that treat it as coordination among individuals mediated by shared artifacts alone. In IA settings, collaborators work within shared spatial–analytic states, carrying out and interpreting analytical actions together in real time. This helps explain why familiar CSCW challenges persist in immersive contexts and why participants emphasized support for maintaining shared alignment.} To \camera{operationalize this understanding of collaboration}, we articulate five empirically grounded design implications (\ref{I1}, ~\ref{I2}, ~\ref{I3}, ~\ref{I4}, ~\ref{I5}), actionable, evidence-based design recommendations inspired from our study’s findings and position them within the extended Time–Space Matrix~\cite{johansen1988groupware} in Table~\ref{tab:cscw_matrix}. Each implication is further mapped to the specific challenges, perceptions, features, and risks it responds to, as summarized in Table~\ref{tab:implications_mapping}.

\begin{figure*}[t]
    \centering
    \includegraphics[width=0.9\linewidth]{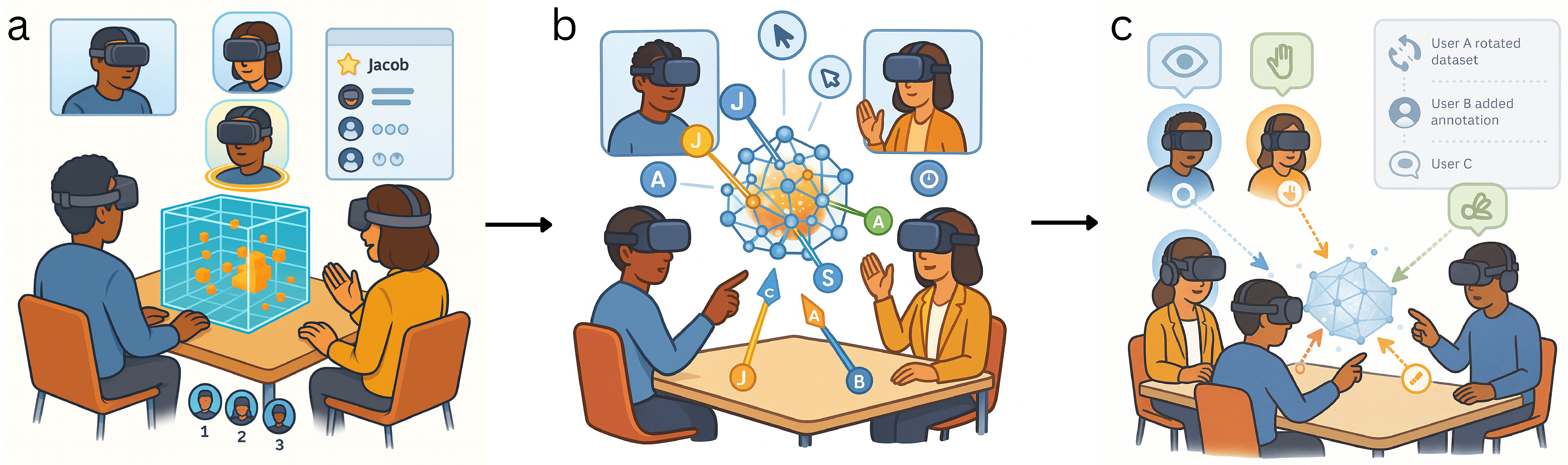}
    \caption{Illustration of how awareness mechanisms may support social translucence in cross-modal CIVA environments. (a) Turn-Taking \& Roles: Active speaker marked, roster shows roles, and avatar queue manages speaking order; (b) Awareness Cues: Telepointers, gaze beams, gestures, and activity badges reveal actions and intentions; (c) Visibility \& Accountability: Glowing marks, action icons, and an activity log ensure transparency and trust. Illustrations are generated using GenAI tools and refined by the authors.}
    \Description{The illustration, set against a clean white background, shows four participants collaborating in immersive analytics across three design perspectives. In the Turn-taking and Role Management panel, two co-located participants wear VR headsets at a shared table, while two remote collaborators appear in holographic panels. A glowing 3D dataset anchors the session. One participant is visually highlighted as the active speaker (floor control), a roster panel shows participant roles (moderator, speaker, observer), and a small queue of avatars illustrates how turn-taking is managed. In the Awareness Cues panel, the same four participants engage around a volumetric network dataset. Telepointers with initials point to nodes, gaze beams extend from headsets to areas of focus, gestures such as pointing or raising a hand are visible, and activity badges (editing, selecting, waiting) float above avatars. These mechanisms make attention and intent visible without overwhelming the shared data. In the Social Translucence panel, participant activity is made transparent and accountable. Each avatar is outlined with a glowing aura showing visibility, intuitive icons (eye, hand, speech bubble) explain actions, and a semi-transparent activity log panel records recent actions (e.g., rotating or annotating the dataset), with dotted lines linking entries to avatars. Together, the three panels illustrate how structured roles, awareness cues, and translucence principles combine to support smooth and trustworthy collaboration in immersive analytics.}
    \label{fig:awareness_framework}
\end{figure*}

%%%%%%%%%%%%%%%%%%%%%%
\subsection{Coordinating Work through Shared Computational History}\label{I1}
%%%%%%%%%%%%%%%%%%%%%%

\begin{figure*}[t]
    \centering
    \includegraphics[width=0.99\linewidth]{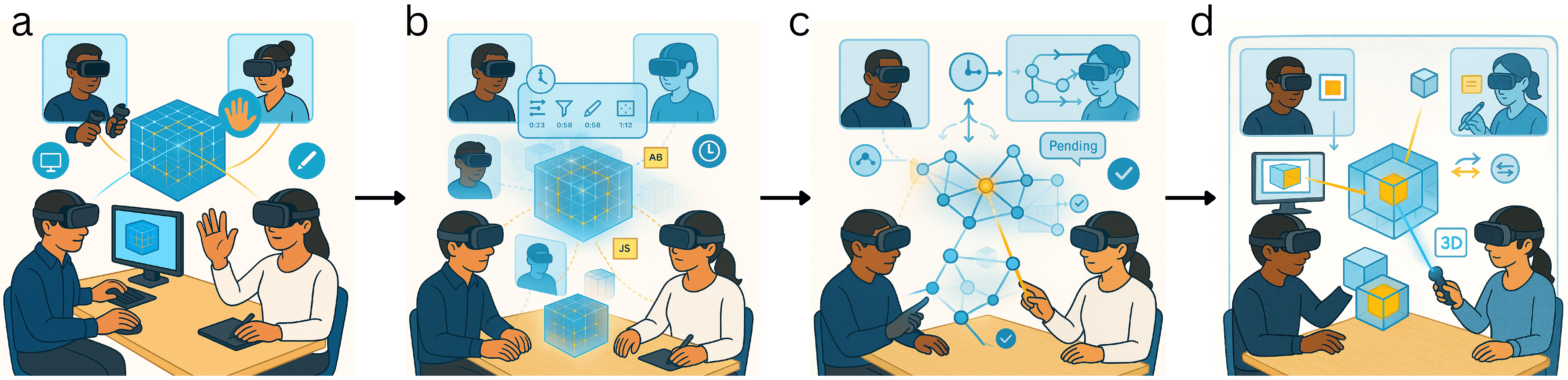}
    \caption{Prospective cross-device workflows for hybrid presence in CIVA; (a) Device-agnostic interaction with desktops, VR controllers, hand-tracking, and stylus on a shared dataset; (b) Session management with timelines, preserved history, and rejoining for continuity and accountability; (c) Synchronization through shared timing, conflict handling, and smooth updates across users; (d) Hybrid presence where co-located VR and remote participants collaborate across 2D–3D interfaces with boundary management and real-time synchronization. Illustrations are generated using GenAI tools and refined by the authors.}
    \Description{Together, the four figures represent how cross-device workflows enable hybrid presence in CIVA. They highlight the structured foundation of device-agnostic interaction, session management, synchronization, and hybrid presence that supports continuity, coordination, and shared awareness across participants and interfaces. (a) Device-Agnostic Interaction. Participants collaborate with desktops, VR controllers, hand-tracking, and a stylus, all seamlessly manipulating the same 3D dataset. Glowing beams converge on the data, symbolizing device-agnostic abstractions and illustrating how diverse inputs maintain a unified shared environment. (b) Session Management. A semi-transparent timeline panel records rotations, filters, annotations, and selections with timestamps. Ghosted versions of the dataset preserve interaction history, dotted lines link participants to past actions, and sticky-note initials indicate authorship. A rejoining avatar fades in to recover preserved context, reflecting continuity, provenance, and accountability across sessions. (c) Synchronization. A central clock represents shared timing, while merging arrows visualize conflict resolution and a background vector clock encodes the order of actions. “Pending” bubbles depict delayed updates that integrate smoothly, and color highlights connect changes to users—supporting coordination, awareness, and consistent shared state.(d) Hybrid Presence. Co-located VR participants and remote collaborators work together on the same dataset across 2D and 3D interfaces. Gestures and controllers manipulate volumetric data, a desktop collaborator edits a linked 2D slice via a sync beam, and a stylus user contributes annotations. A semi-transparent boundary frame bridges the physical table and remote panels, symbolizing hybrid presence, boundary negotiation, and shared awareness.}
    \label{fig:sync_framework}
\end{figure*}

Figure~\ref{fig:provenance_swimlanes} illustrates how multi-user branching, collaboration, and merging can be captured through provenance mechanisms, converging on a design direction focused on preserving analytical histories across distributed teams.

The articulation work framework~\cite{schmidt1992taking} emphasizes that collaborative systems must continuously support coordination to align distributed efforts toward shared goals. Our findings reveal a breakdown in \textit{representational state propagation}~\cite{hutchins1995cognition}, the transfer of information and context across participants and over time. Participants’ recurring demands for ``snapshots'', ``recordings'', and ``who did what when'' tracking indicate failures in \textit{boundary object maintenance}~\cite{star1989institutional}, where shared meaning is lost across perspectives and temporal boundaries. To address these challenges, provenance must move beyond simple interaction logs to capture \textit{interaction intent}~\cite{ragan2015characterizing}, recording not only user actions but also their underlying analytical motivations.  

A temporal provenance graph can represent analytical states (e.g., camera views, filters, selections, annotations) as nodes and encode transformations with parameters and user attribution as edges. This structure supports \textit{analysis branching and merging}~\cite{brehmer2013multi}, enabling collaborators to explore alternative pathways while maintaining continuity with shared analytical narratives. Implementing such provenance requires hybrid client–server infrastructures that propagate lightweight deltas in real time while persisting full snapshots for reproducibility. Server-side hosting centralizes computation and synchronization, avoiding redundant local copies, ensuring consistent state updates, and supporting lightweight clients.

%%%%%%%%%%%%%%%%%%
\subsection{Enhancing Awareness through Social Translucence}\label{I2}
%%%%%%%%%%%%%%%%%%

Figure~\ref{fig:awareness_framework} illustrates how awareness mechanisms support social translucence in cross-modal CIVA environments, pointing to a design recommendation for making actions visible, interpretable, and accountable across heterogeneous devices.

The social translucence framework~\cite{erickson2000social} requires participants' actions to be visible, interpretable, and accountable for effective collaboration. Our findings reveal that current immersive analytics systems lack sufficient awareness mechanisms, causing coordination failures and lapses in \textit{situation awareness}~\cite{endsley1995measurement}.Systems must implement awareness mechanisms, including real-time telepointers, selection highlighting with user attribution, optional gaze and gesture indicators, and activity badges showing manipulation states. These features support \textit{workspace awareness}~\cite{gutwin2002descriptive} by helping collaborators track locations, actions, and intentions. Cross-modal awareness translation presents implementation challenges, particularl 
where desktop users coordinate with VR participants~\cite{gottsacker2025decoupled}, requiring approaches that maintain awareness across different temporal participation rhythms and interface modalities~\cite{lee2021xr,irlitti2023volumetric}.

Large-group collaboration demands explicit orchestration through floor control mechanisms: presenter locks, raise-hand queues, and speaker identification with spatial indicators. These \textit{structured interaction protocols}~\cite{monk2003common} address turn-taking challenges that scale non-linearly with group size~\cite{arrow2000small}. Practice-informed patterns include participant rosters, role indicators, speaker beacons, request queues, and per-user data layers that enable individualized perspectives while sustaining collaborative participation~\cite{wong2024practice}. Awareness cues must remain peripheral to primary analytical tasks, informing without overwhelming scientific sensemaking.

\begin{figure*}[t]
    \centering
    \includegraphics[width=0.82\linewidth]{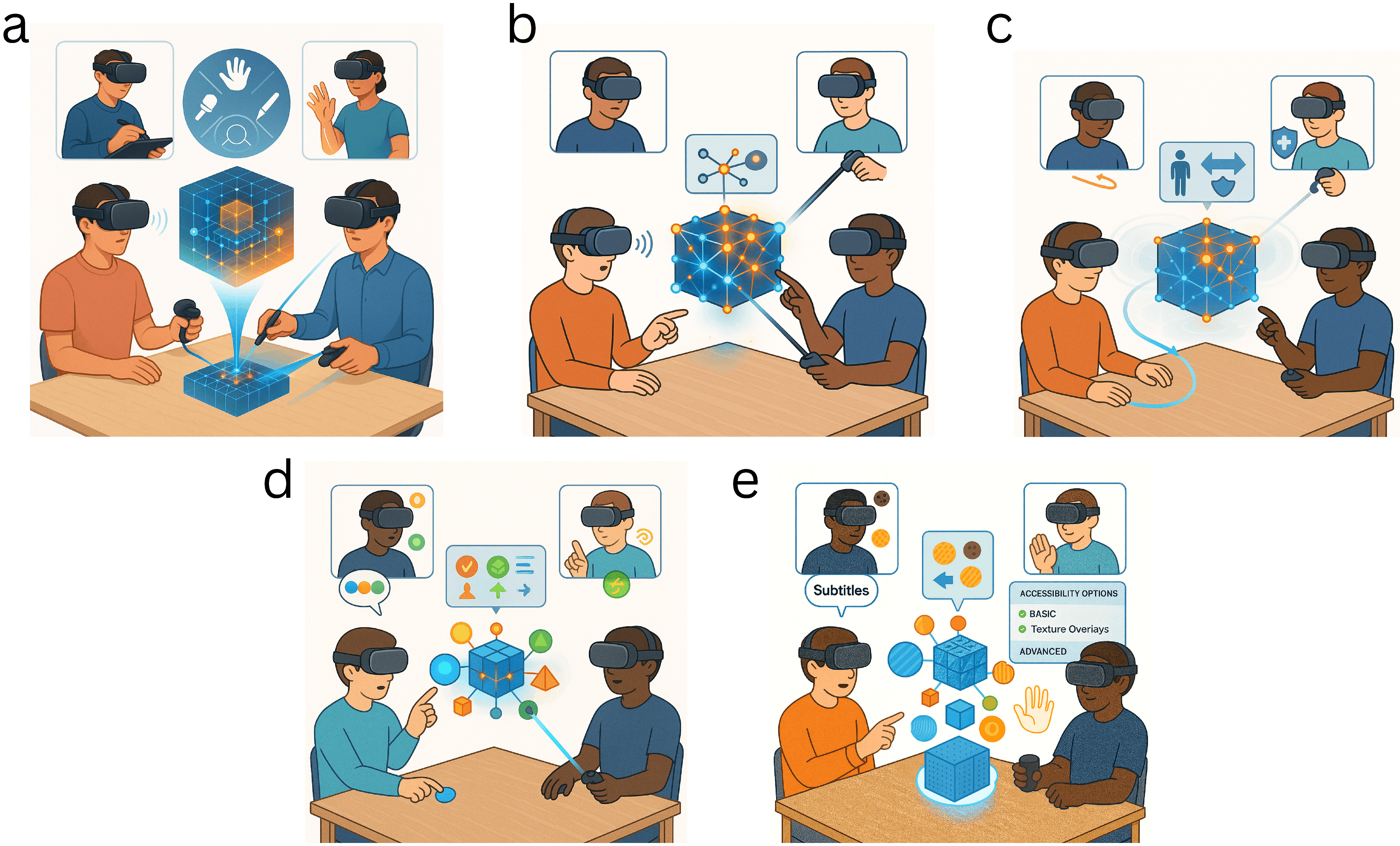}
    \caption{Illustration of accessibility integration in hybrid CIVA, shown as an upward pathway culminating in Inclusive Co-Design. Each sub-panel demonstrates a progressive layer of support: (a) Universal Interaction Protocols ensure equal participation across modalities (speech, gesture, stylus, controllers); (b) Multimodal Mutual Disambiguation resolves ambiguity by aligning cues (speech, gaze, gesture, controller rays) on the same dataset region; (c) Embodiment \& Comfort harmonizes locomotion and viewpoint alignment with adaptive comfort dashboards; (d) Redundant Encoding reinforces clarity through multiple perceptual channels (color, shape, motion, legends); and (e) Progressive Accessibility Enhancement integrates subtitles, textures, adaptive devices, and simplified gestures, showing how immersive analytics adapt to diverse user needs through inclusive design. Illustrations are generated using GenAI tools and refined by the authors.}
    \Description{This collage illustrates the progressive integration of accessibility features in CIVA, framed along an upward pathway that culminates in Inclusive Co-Design. Each stage builds on the previous, moving from universal participation protocols toward increasingly adaptive and inclusive mechanisms that ensure all collaborators can meaningfully engage with complex 3D data.
    (a) Universal Interaction Protocols: Four collaborators—two co-located at a table and two remote in panels—interact with a shared cube-grid dataset using different modalities: speech, stylus, gestures, and VR controllers. A unifying protocol ring highlights parity across modalities, ensuring no single input dominates. The scene emphasizes fairness, inclusivity, and social translucence, representing a foundational step toward equitable participation.
    (b) Multimodal Mutual Disambiguation: Participants’ diverse cues—speech bubbles, gaze beams, pointing gestures, and controller rays—converge on the same highlighted cube cluster within the dataset. The convergence glows to indicate collective agreement, demonstrating how overlapping input streams reinforce shared clarity and prevent misinterpretation. This stage highlights the principle of ambiguity resolution through multimodal synergy.
    
    (c) Embodiment & Comfort: The illustration shows collaborators navigating the immersive space using different locomotion modes: teleportation arcs, joystick path lines, viewpoint rotation indicators, and adaptive comfort shields. A floating comfort dashboard displays posture, alignment, and motion safety cues, while faint viewpoint outlines help participants stay synchronized. The relaxed body language and natural gestures emphasize inclusivity through ergonomic support and adaptive locomotion.
    (d) Redundant Encoding: The cube-grid dataset is encoded with multiple perceptual channels: bold colors (blue, orange, green), geometric shapes (spheres, cubes, pyramids), and motion cues (pulsing, trails). Each participant attends to a different cue yet converges on the same cluster. A shared legend clarifies mappings, while recognition icons above avatars reinforce consistent interpretation. This stage emphasizes accessibility through redundancy, ensuring clarity for participants with different perceptual strengths.
    (e) Progressive Accessibility Enhancement: Accessibility features are layered to adapt to diverse needs: subtitles appear below a speaking avatar, clusters are distinguished with patterned textures for colorblind accessibility, an adaptive controller grip glows with a comfort icon, and simplified gestures are highlighted. A floating Accessibility Options panel lists active features, including subtitles, overlays, and locomotion aids. This final stage represents progressive inclusivity, showing how immersive analytics systems can flexibly adapt to individual abilities and collectively support Inclusive Co-Design.}
    \label{fig:accessibility_framework}
\end{figure*}

%%%%%%%%%%%%%%%%%%%%%%
\subsection{Synchronizing Cross-Device Workflows for Hybrid Presence}\label{I3}
%%%%%%%%%%%%%%%%%%%%%%

Figure~\ref{fig:sync_framework} illustrates prospective cross-device workflows for hybrid presence in CIVA, underscoring a design recommendation for synchronization protocols that sustain coherent collaboration across heterogeneous devices and hybrid presence contexts.

Our findings reveal critical coordination challenges when scientists maintain a consistent analytical state across heterogeneous devices such as desktop workstations, VR headsets, and AR displays~\cite{rekimoto1997pick}. This creates \textit{distributed embodiment}, where collaborators engage through different sensorimotor interfaces but require coherent shared context. Unlike \textit{embodied interaction}, which assumes a single interface~\cite{dourish2001action}, distributed embodiment requires that analytical operations, such as data selection, filtering, and annotation, remain semantically equivalent across modalities. Systems must implement \textit{device-agnostic interaction abstractions} that preserve analytical intent across platforms, supported by \textit{cross-device capability negotiation}, adapting functions to each participant’s technological context~\cite{radle2015spatially}. Session management must further maintain \textit{embodied session state}~\cite{pierce2008infrastructure} to support continuity in CIVA environments. This entails preserving spatial relationships, interaction histories, and shared analytical context as participants transition across devices and modes of interaction.

%Session management must further maintain \textit{embodied session state}~\cite{pierce2008infrastructure}, preserving spatial relationships, interaction histories, and collaborative context as participants transition between devices.

Effective synchronization requires handling \textit{temporal coupling variations}~\cite{grudin1994groupware}, where scientists engage asynchronously or at different paces,
through mechanisms such as vector clocks that preserve causal ordering 
~\cite{lamport2019time}. Hybrid presence scenarios demand sophisticated \textit{boundary management}~\cite{benford1998understanding} to achieve functional parity despite varying spatial affordances. Desktop participants exploring 2D projections must coordinate seamlessly with VR collaborators manipulating 3D data volumes, necessitating \textit{asymmetric collaborative interfaces}~\cite{billinghurst2002collaborative}, that deliver equivalent analytical capabilities through modality-appropriate interaction patterns. These requirements establish cross-device synchronization as essential infrastructure for CIVA, enabling scientific teams to integrate diverse technologies while sustaining coherent workflows.

%%%%%%%%%%%%%%%%%%%%%%
\subsection{Integrating Accessibility into Collaborative Infrastructures}\label{I4}
%%%%%%%%%%%%%%%%%%%%%%

\begin{figure*}[t]
  \centering
  \includegraphics[width=0.8\textwidth]{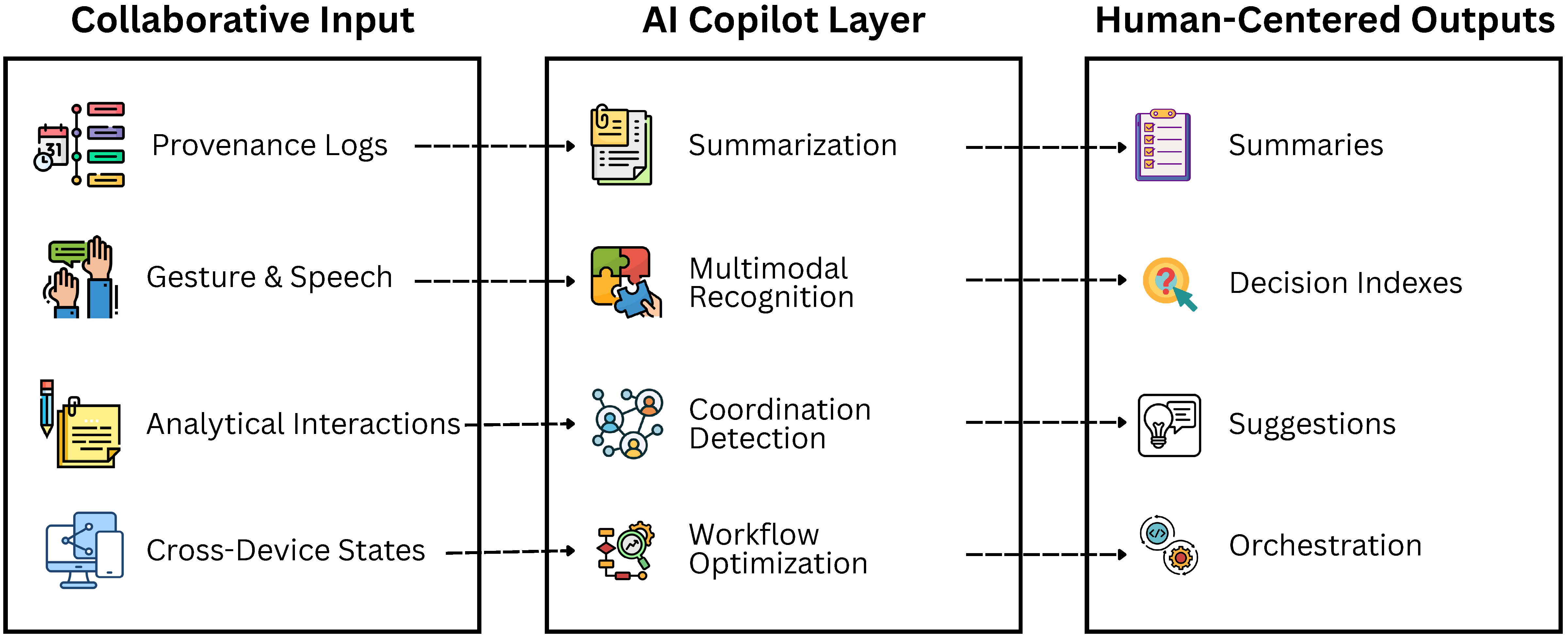}
  \caption{Proposed AI-mediated coordination framework for CIVA.}
  \Description{The framework illustrates how Collaborative Inputs (e.g., provenance logs, multimodal signals, analytical interactions, and cross-device states) flow into the AI Copilot Layer, where functions such as provenance summarization, multimodal recognition, coordination detection, and workflow optimization transform raw activity into structured insights. These are then surfaced as Human-Centered Outputs, including session summaries, decision indexes, analytical suggestions, and orchestration prompts. The model emphasizes the core principle of human analytical ownership augmented by AI-driven coordination amplification, highlighting how AI acts as an intermediary that enhances—but does not replace—human judgment in distributed immersive settings.}
  \label{fig:implication5_ai}
\end{figure*}

Figure~\ref{fig:accessibility_framework} illustrates our proposed layered accessibility framework, progressing from universal interaction protocols to inclusive co-design and integrating accessibility into the collaborative foundations of CIVA.

Our findings reveal that when immersive systems fail to accommodate users with disabilities, these individual barriers create cascading effects that exclude entire teams from adopting analytical tools, a phenomenon termed \textit{collaborative exclusion cascades}~\cite{ladner2015design}. Teams working with diverse sensory, motor, and cognitive abilities require \textit{ability-diverse collaboration} frameworks that maintain analytical effectiveness for all participants without disadvantaging any team member~\cite{mankoff2010disability}. Research demonstrates that these exclusionary patterns stem from inadequate system design rather than user limitations~\cite{xiao2024systematic}, emphasizing the need for collaborative infrastructures that proactively address immersive technology barriers~\cite{creed2024inclusive} at scale through comprehensive accessibility strategies for AR and VR environments~\cite{dudley2023inclusive}.

Successful collaborative accessibility requires several integrated mechanisms.
\textit{Universal interaction protocols} must operate consistently across contexts and abilities~\cite{wobbrock2011ability}, supported by \textit{mutual disambiguation} mechanisms that let collaborators clarify and complement one another’s actions across multiple input channels~\cite{oviatt1999ten}. Shared immersive environments should incorporate \textit{adaptive locomotion algorithms} that adjust movement to individual comfort~\cite{laviola2000discussion} and \textit{comfort-preserving viewpoint synchronization} to mitigate motion sickness~\cite{fernandes2016combating}. Information should use \textit{redundant encoding}—via color, texture, or motion—to support consistent interpretation across perceptual differences~\cite{ware2019information}. Systems should follow \textit{progressive accessibility enhancement}, offering baseline accessibility features with optional extensions for specific needs~\cite{froehlich2007barrier}, while avoiding \textit{accessibility–performance tradeoffs} that degrade overall effectiveness~\cite{ardito2020analysing}. Evidence from inclusive \textit{co-design}~\cite{flacke2025inclusive} shows that embedding accessibility from the outset establishes it as a foundational requirement for CIVA.

%%%%%%%%%%%%%%%%%%%%%%
\subsection{Orchestrating Collaboration through AI-Mediated Coordination}\label{I5}
%%%%%%%%%%%%%%%%%%%%%%

Figure~\ref{fig:implication5_ai} depicts how AI copilots enhance collaborative intelligence through improved recall, coordination cues, and workflow guidance while preserving full human analytical agency, offering a design implication for AI-supported coordination across teams.

The complexity of CIVA creates opportunities for AI to reduce coordination overhead while preserving human analytical agency.  Our participants expressed interest in automated session summarization, gesture and speech recognition, and intelligent suggestions for next analytical steps, reflecting broader trends in \textit{human–AI collaboration}~\cite{amershi2019guidelines} where AI systems augment rather than replace human decision-making. AI should function as a \textit{copilot} that supports articulation work, the ongoing coordination activities required for collaborative scientific analysis, by automatically summarizing provenance logs, indexing key analytical decisions, and proposing next steps for team consideration. Recent advances demonstrate this feasibility: LLMs, VLMs can effectively summarize computational workflows through \textit{provenance summarization}\cite{boufford2024computational}, while AI agents functioning as \textit{spatial collaborators}~\cite{fernandez2025augmenting} show promise for gesture- and speech-based assistance in multimodal scientific environments. Enterprise tools such as Microsoft Copilot~\cite{copilot} and large multimodal models such as GPT~\cite{gpt5} and Gemini~\cite{gemini} provide technical foundations for provenance analysis, multimodal interaction understanding, and cross-device coordination support.

Trustworthy AI-mediated coordination must adhere to \textit{human-centered AI} principles~\cite{shneiderman2020human}, emphasizing transparency in suggestion generation, explainability of AI reasoning, and user control over analytical decisions. The concept of \textit{mixed-focus collaboration}~\cite{cheng2025transitioning} demonstrates how control can flexibly shift between humans and AI depending on analytical context, with AI handling routine coordination tasks while humans maintain ownership of scientific insights and conclusions. This human-AI partnership model addresses the fundamental challenge of scaling CIVA: as team sizes and data complexity increase, coordination demands overwhelm analytical productivity, yet this approach enables CIVA to scale while preserving the scientific rigor and human insight essential for high-stakes research decisions.

%%%%%%%%%%%%%%%%%%%%%%
\section{Limitations and Future Work}\label{limitation}
%%%%%%%%%%%%%%%%%%%%%% 
Our recruitment strategy relied on remote interviews with domain experts working with high-dimensional data, which may introduce sample skew toward experienced researchers. Though sessions were remote and participants could not interact with immersive systems directly, their expertise in data analysis, visualization, and immersive technologies provided a strong basis for articulating collaboration needs in CIVA. To mitigate potential priming effects, all participants viewed the same design-probe video and were asked open-ended questions. We acknowledge our positionality as researchers in IA and its influence on interpretation. \camera{Throughout the findings, we explicitly discuss how participant expertise, team structures, and methodological choices shaped the collaborative dynamics identified in our analysis. This reflexive framing informs both the interpretation of findings and the development of design implications.}

As CIVA platforms rely on advanced hardware and distributed infrastructure, future work should examine sustainable development models, including open-source ecosystems and cross-institutional partnerships to support adoption in under-resourced environments such as non-profit research centers and state-run medical facilities. Building on the design insights identified in this study, future work can develop CIVA systems for structured evaluations of collaboration quality, task performance, and accessibility. Comparative studies across 2D, immersive, and hybrid environments, supported by shared benchmark datasets and standardized evaluation frameworks, will enable systematic assessment of collaborative workflows. Incorporating perspectives beyond visualization and immersive analytics will further enrich understanding of CIVA’s role across diverse scientific domains.

%%%%%%%%%%%%%%%%%%%%%%
\section{Conclusion}\label{conclusion}
%%%%%%%%%%%%%%%%%%%%%% 

Our qualitative study characterizes the emerging landscape of CIVA through four interconnected themes spanning current workflow challenges, perceptions of adoption, prospective system features, and anticipated usability and ethical risks. We find that scientific teams continue to work within fragmented, tool-incompatible ecosystems that limit real-time shared analysis, yet they express cautious optimism toward adopting CIVA when it aligns with established computational workflows and minimizes setup friction. Participants envision immersive environments that support synchronized multi-user interaction, cross-device hybrid presence, rich collaborative provenance, and AI-mediated orchestration, while anticipating constraints related to accessibility, data governance, interaction precision, comfort, and equitable participation. Overall, our findings establish foundational guidance for future CIVA systems, positioning immersive collaboration as a prospective and complementary infrastructure for advancing team-based reasoning over high-dimensional scientific data.

\begin{acks}
We extend our thanks to the anonymous reviewers for their insightful comments and to the interview participants for their valuable contributions.
This work was partially supported by the U.S. National Science Foundation under grant CNS-2235049.
Certain commercial equipment, instruments, or materials (or suppliers, or software, etc) are identified in this paper to foster understanding. Such identification does not imply recommendation or endorsement by the National Institute of Standards and Technology, nor does it imply that the materials or equipment identified are necessarily the best available for the purpose. 
\end{acks}

\newpage
\bibliographystyle{ACM-Reference-Format}
\bibliography{references}

%\clearpage
\appendix
\section{Video Links}\label{Video_Link}
\begin{table}[h]
\centering
\small
\begin{tabular}{p{2.8cm}|p{4.5cm}}
\textbf{Video Demo Title} & \textbf{Video Link} \\
\hline
ParaView XR Interface & \href{https://youtu.be/O3t14Vz59I4?si=EraiCvQuMAIZ5tEY}{https://youtu.be/O3t14Vz59I4} \\
\hline
ImAxesGEO & \href{https://youtu.be/kyQ_GWoSy_0?si=XwsmiEuYne2pGBEb}{https://youtu.be/kyQ\_GWoSy\_0} \\
\hline
Uplift & \href{https://youtu.be/JrH2dVuxa1I?si=Z7rfGRej_GV4vDFW}{https://youtu.be/JrH2dVuxa1I} \\
\hline
EarthGraph & \href{https://youtu.be/s2B5DdPc8aI}{https://youtu.be/s2B5DdPc8aI} \\
\hline
Embodied Axes & \href{https://youtu.be/p19ub_pGN5U?si=WGnKi1NZZtSmR8lA}{https://youtu.be/p19ub\_pGN5U} \\
\hline
FIESTA & \href{https://doi.org/10.1145/3343055.3360746}{https://doi.org/10.1145/3343055.3360746} \\
\hline
Data Visualization in VR & \href{https://youtu.be/OOOVEpgIIJI?si=x0h8wmn1U2eanQEw}{https://youtu.be/OOOVEpgIIJI} \\
\hline
VR-Assisted Microscopy Data Visualization & \href{https://youtu.be/ajdNKnAHFMw?si=KRg2tMU-tD9y-Lka}{https://youtu.be/ajdNKnAHFMw} \\
\hline
HammerheadVR & \href{https://youtu.be/sUvDAieci1s?si=WRIwjbiVopiicrDL}{https://youtu.be/sUvDAieci1s} \\
\hline
See Through Brains & \href{https://youtu.be/c-NMfp13Uug?si=AfOpaOc3a7PfrJT3}{https://youtu.be/c-NMfp13Uug} \\
\end{tabular}
\caption{Video demos with corresponding publicly available links.}
\label{tab:video-demos-links}
\end{table}

\end{document}